\documentclass[utf8]{frontiersFPHY} 
\pdfoutput=1

\usepackage{geometry,amsmath,amsfonts}
\usepackage{slashed}
\usepackage{epsfig}
\usepackage{latexsym}
\usepackage{graphicx}
\usepackage[justification=RaggedRight]{caption}
\usepackage{amssymb}
\usepackage{stfloats}
\usepackage{color}
\usepackage{multirow}
\usepackage{color}
\usepackage{rotating}
\usepackage{ifthen}
\def\firstAuthorLast{G. Arcadi and A. Djouadi}
\def\Authors{Giorgio Arcadi$^1$ and Abdelhak Djouadi$^{2,3}$ }
\def\Address{\small
$^1$ Dpt. MIFT, %di Scienze Matematiche e Informatiche, Scienze Fisiche e Scienze della Terra, 
Universita degli Studi di Messina, Via Ferdinando Stagno d'Alcontres 31, I-98166 Messina, Italy.\\
$^2$ CAFPE and Departamento de F\'isica Te\'orica y del Cosmos, Universidad de Granada, E-18071 Granada, Spain.\\
$^3$ Laboratory of High Energy and Comp. Physics, R\"avala pst. 10, 10143 Tallinn, Estonia.
}
\def\corrAuthor{Giorgio Arcadi} 
\def\corrEmail{giorgio.arcadi@unime.it}

\begin{document}
\onecolumn
\firstpage{1}

\title[Dark Matter and the $\mathbf{M_W}$ and $\mathbf{(g-2)_\mu}$ anomalies]{
A model for fermionic dark matter addressing both the CDF $\mathbf{M_W}$ and the $\mathbf{(g-2)_\mu}$ anomalies} 

\author[\firstAuthorLast ]{\Authors} %This field will be automatically populated
\address{} %This field will be automatically populated
\correspondance{} %This field will be automatically populated

\extraAuth{}

\maketitle

\begin{abstract}

\section{}
We explore a simple and renormalizable model which incorporates a new stable fermion that accounts for the dark matter in the universe and which, at the same time,  provides an interpretation of two recent measurements that deviate from the expectation in the Standard Model: the muon anomalous magnetic moment and the mass of the W-boson recently performed by the CDF collaboration. The model has a fermionic sector that involves a singlet and a doublet fields and in which the lightest state is the DM and interacts mainly through the Higgs portal. Two realizations of such a possibility are considered: one in which the Higgs sector is minimal and another in which it is extended to contain two doublet fields. 
\end{abstract}

\section{Introduction} 

There is a large consensus that the Standard Model (SM) of particle physics, although chiefly confirmed by the recent campaign of direct and indirect searches performed at the CERN LHC \cite{ATLAS-new,CMS-new}, cannot be the ultimate theory and should only be valid at currently explored energies. One of the main reasons is that it does not involve an electrically neutral, weakly interacting and massive particle (WIMP) that could account for the dark matter that apparently forms 25\% of the energy budget of the Universe \cite{Planck:2018vyg}. New physics beyond the SM that incorporates such a particle is thus highly expected \cite{Bertone:2004pz,Arcadi:2017kky}. Most interesting are the scenarios in which this WIMP interacts mainly or exclusively through the Higgs sector of the theory, the so-called Higgs portal models; see Ref.~\cite{Arcadi:2019lka} for a recent review.

The need for new physics beyond the SM received recently a further boost with some unexpected experimental results performed at Fermilab and which cannot be interpreted in the strict context of the model. The most surprising one is a new and more precise determination of the mass of the $W$ boson performed by the CDF collaboration \cite{CDF:2022hxs} 
\begin{equation} 
M_W = 80.4335 \pm 0.0094~{\rm GeV}\, ,  
\end{equation}
which not only deviates by about $7\sigma$ with respect to the SM value but is also in conflict with similar measurements performed at other colliders such as LEP and LHC \cite{ParticleDataGroup:2020ssz}.  Another result which was less surprising is the latest Fermilab measurement of the anomalous magnetic magnetic moment of the muon, $a_\mu= \frac12 (g-2)_\mu$, which was found to be 
\cite{Muong-2:2021ojo}
\begin{equation} \label{eq:4sigma}
    a_\mu^{\rm EXP}= (116 592 061 \pm 41)\times 10^{-11} \, ,
\end{equation}
confirming and magnifying the long standing $(g\!-\!2)_\mu$ anomaly \cite{Muong-2:2006rrc}, as the deviation from the consensus theory prediction in the SM \cite{Aoyama:2020ynm} is now $4.2\sigma$, $\Delta a_\mu\!= \!a_\mu^{\rm EXP} \!- \!a_\mu^{\rm SM} \! = \! ( 251 \! \pm \! 59) \! \times \! 10^{-11}$. Again, there is an ongoing debate about the validity of the SM prediction and the possibility that the discrepancy could partly be due to unknown uncertainties, as suggested by a recent conflicting lattice calculation \cite{Borsanyi:2020mff}, should not be overlooked.\vspace*{-1mm} 
 
Despite that these two results are still controversial and require further experimental and theoretical scrutiny, it is tempting to interpret them as the first of the long awaited hints of new physics beyond the SM.  Nevertheless, one should at least try to relate the two anomalies and explore the possibility of embedding them into model extensions that address also some important shortcomings of the SM, in particular, those which include a viable DM candidate. This is what we attempt and explore in the present work.\vspace*{-1mm}   

We consider a rather simple particle physics scenario dubbed singlet-doublet model \cite{Cohen:2011ec,Cheung:2013dua,Calibbi:2015nha} in which the DM is the lightest electrically neutral state of a new fermionic sector consisting of an admixture of SU(2) singlet and doublet fields. The new fermions obey a discrete symmetry which forces the DM to be stable and to interact with the SM particle mainly through the Higgs sector. This singlet-doublet model is thus an economical and renormalizable realization of a Higgs portal to a fermionic DM \cite{Arcadi:2019lka}. The extension of the SM spectrum with these fermions, charged under the SU(2) group, implies new contributions that could provide a theory interpretation of the $M_W^{\rm CDF}$ measurement.\footnote{Other extensions, like for instance, the ones discussed in Refs.~\cite{Arcadi:2021zdk,Arcadi:2022dmt,Arcadi:2022lpp} in which the DM is an isosinglet fermion and the Higgs sector is enlarged to contain two doublets and a singlet pseudoscalar Higgs fields, can also achieve this goal.}\vspace*{-1mm} 

Two realizations of this possibility will be considered. In the first and minimal one, the DM state interacts with the SM via the single SM Higgs particle \cite{Djouadi:2005gi} with a mass of 125 GeV observed at the LHC. We will show that, despite that it could lead to a correct DM cosmological relic abundance assuming the freeze-out paradigm \cite{Bertone:2004pz,Arcadi:2017kky}, the model is severely constrained, in particular, by direct DM detection in astroparticle physics experiments which exclude  most of its parameter space. In addition, such a minimal extension, while it can indeed address the $M_W$ anomaly, does not explain the  $(g-2)_\mu$ value. We therefore also consider an extension of the model in which the Higgs sector is enlarged and includes two Higgs doublets fields to break the electroweak symmetry. This two Higgs-doublet model (2HDM) \cite{Branco:2011iw} allows to evade the constraints from DM direct searches while leading to a correct DM relic density and, at the same time, to address both the $M_W$ and $(g-2)_\mu$ anomalies via the new contributions of the richer Higgs sector.\vspace{-1mm} 

The paper is organized as follows:  we introduce the fermionic singlet-doublet model with the minimal SM Higgs sector in the next section and the 2HDM 
extension in section 3. In both sections, we discuss the impact on DM phenomenology and attempt to explain the values of $M_W^{\rm CDF}$ and eventually $(g-2)_\mu$. A short conclusion is given in section 4.

\section{The singlet-doublet model with a SM-like Higgs sector}

\subsection{The theoretical set-up} 

The so-called fermionic singlet-doublet model~\cite{Cohen:2011ec,Cheung:2013dua,Calibbi:2015nha} (see also Refs.~\cite{Yaguna:2015mva,Arcadi:2018pfo,Arcadi:2019lka}) is one of the most minimal ultraviolet-complete realizations of the Higgs-portal framework for dark matter, enabling the possibility of renormalizable interactions between a fermionic DM candidate and the SM Higgs doublet field. In this scenario, the spectrum of the SM is extended by two additional ${\rm SU(2)_L}$ doublets and one singlet fermionic fields
\begin{equation}
D_L=\left( \begin{array}{c} N_L \\ E_L \end{array} \right), \,\,\,\,\,
D_R=\left( \begin{array}{c} -E_R \\ N_R \end{array} \right), \,\,\, S \; , 
\end{equation}
which are described by the following Lagrangian
\begin{equation}
\label{eq:SD_lagrangian}
\mathcal{L}=-\frac{1}{2} m_S S^{2}-m_D D_L D_R -y_1 D_L \Phi S-y_2 D_R \widetilde{\Phi} S+\mbox{h.c.},
\end{equation}
with the implicit assumption that the new states are odd under a $Z_2$ symmetry that forbids mixing with the SM fermions. $\Phi$ is the SM Higgs doublet which, in the unitary gauge, is
\begin{equation}
    \Phi=\frac{1}{\sqrt{2}}\left(
    \begin{array}{c}         0 \\  v+H \end{array} 
    \right) \ ,  \quad v \simeq 246 \ {\rm GeV}     \, . 
\end{equation}
After electroweak symmetry breaking, mixing occurs between the electrically neutral components of the new fermionic fields. The mass eigenstates will be assumed to be three Majorana fermions\footnote{The possibility of Dirac fermions was proposed in Ref.~\cite{Yaguna:2015mva} and leads to a similar picture compared to the Majorana case}.\vspace*{-2mm} whose masses are obtained by diagonalizing the mass matrix
\begin{equation}
\label{eq:SD_mass_matrix}
{\mathcal M} =\left(
\begin{array}{ccc}
m_S & {y_1 v}/{\sqrt{2}}~~ & {y_2 v}/{\sqrt{2}}~~ \\
{y_1 v}/{\sqrt{2}}~~ & 0 & m_D \\ {y_2 v}/{\sqrt{2}}~~ & m_D & 0
\end{array}
\right)\, . 
\end{equation}
The mass eigenstates, using the unitary $3\!\times\! 3$ matrix $U$ diagonalizing  ${\mathcal M}$, are defined as
\begin{equation}
    \chi_i=S U_{i1}+N_L U_{i2}+N_R U_{i3}\, , 
\end{equation}
with, by convention, $m_{\chi_1} \!< \! m_{\chi_2} \! < \! m_{\chi_3}$.  The electrically charged components of the new fermionic fields form instead a Dirac fermion, which we denote $\psi^{\pm}$ and with a mass $m_{\psi^{\pm}} \simeq m_D$. If $m_{\chi_1}\!<\! m_{\psi^{\pm}}$, the lightest Majorana fermion will be the DM candidate as, by virtue of the $Z_2$ discrete symmetry, it will be absolutely stable. 

In the physical basis, the interaction Lagrangian of the new fermions reads \cite{Arcadi:2017kky}
\begin{align}
\label{eq:physical_SD}
\mathcal{L}&=\bar \chi_i \gamma^\mu \left(g_{Z \chi_i \chi_j}^V -g_{Z \chi_i \chi_j}^A \gamma_5\right) \chi_j Z_\mu+\bar \psi^{-}\gamma^\mu \left(g_{W^{\mp} \psi^{\pm} \chi_i}^V-g_{W^{\mp} \psi^{\pm} \chi_i}^A \gamma_5 \right) W^{-}_\mu \chi_i \nonumber\\
& \!-\!e \bar \psi^- \gamma^\mu \psi^- A_\mu \!-\!\frac{g}{2 \cos^2\theta_W}(1\!-\! 2 \sin^2\theta_W) \bar \psi^- \gamma^\mu \psi^- Z_\mu \!+\! g_{H \chi_i \chi_j}H \bar \chi_i \chi_j +\mbox{h.c.}, 
\end{align}
with $g$ the ${\rm SU(2)_L}$ gauge coupling and $\cos^2\theta_W = 1- \sin^2\theta_W= M_W^2/M_Z^2$. The couplings of the new fermions with the gauge and Higgs bosons can be written, in terms of the elements of the mixing matrix $U$, as   
\begin{align}
    & g_{H \chi_i \chi_j}=\frac{1}{\sqrt{2}}\left(y_1 U_{i2}^{*}U_{j1}^{*}+y_2 U_{j2}^{*}U_{i1}^{*}\right)\, , \quad  g^{V/A}_{W^\mp \psi^{\pm} N_i}=\frac{g}{2\sqrt{2}}\left(U_{i3} \mp U_{i2}^{*}\right) \, , \nonumber \\
    & g^{V/A}_{Z \chi_i \chi_j}=c_{Z \chi_i \chi_j} \mp c^{*}_{Z \chi_i \chi_j},\quad c_{Z \chi_i \chi_j} =\frac{g}{4\cos\theta_W}\left(U_{i3}U_{j3}^{*}- U_{i2}U_{j2}^{*}\right) \, .
\end{align}
From the equations above, one notices in particular that given its Majorana nature, the DM couples in pairs with the $Z$ boson only via the vector-axial interaction; there are also couplings to the $W$ boson. The model is thus, not strictly of the Higgs-portal type and this will have an impact on the phenomenology as will be seen shortly. Following Ref.~\cite{Calibbi:2015nha}, we will trade the parameters $y_1,y_2$ with a single coupling $y$ and a mixing angle $\theta$
\begin{equation}
y_1=y\cos\theta,\,\,\,\,\,y_2=y \sin\theta.
\end{equation} 
With these elements one can start discussing the phenomenology of the model and, in particular, the way it addresses the DM issue and the CDF measurement of $M_W$. 

\subsection{The DM relic density and constraints from direct detection}

In order to be a viable DM candidate, the lightest Majorana fermion should have a primordial abundance which is compatible with the measurement $\Omega_{\rm DM}h^2 \approx 0.12 \pm 0.0012$ performed by the PLANCK experiment \cite{Planck:2018vyg}. Throughout this work, we will assume that the DM relic density is accounted for in the standard thermal freeze-out paradigm in which it is related to a thermally averaged annihilation cross section of the order of $\langle \sigma v \rangle \propto 10^{-26}{\mbox{cm}}^3 {\mbox{s}}^{-1}$ \cite{Bertone:2004pz,Arcadi:2017kky}. In our singlet-doublet model, the DM annihilates mostly into SM fermion pairs via $s$-channel exchange of the $H$ and $Z$ bosons and, for larger DM masses, into $WW,ZZ$ and $Zh$ final states. The latter channels occur not only through $Z$ and $H$ boson exchange but also through $t$-channel exchange of the new fermions.

Note that if the DM is very close in mass to some of its fermionic partners, coannihilation processes involving the DM and these fermions or these fermions themselves, come as a supplement to DM annihilation and could, in any case, provide the correct relic density. In order to determine it with a sufficient accuracy and match it with the PLANCK value, we have implemented the model into the numerical package micrOMEGAs \cite{Belanger:2001fz,Belanger:2007zz} which includes all (co)annihilation channels and all relevant effects.

There are other constraints on the DM mass and couplings beyond the one from the relic density and the strongest one comes from direct detection  in astroparticle experiments, i.e. in elastic scattering of the DM with nuclei. Our singlet-doublet DM model features both the spin-independent (SI) and the  spin-dependent (SD) interactions. The former are due to the interaction of the DM with the Higgs boson and are described by the following DM-nucleon scattering cross section (for simplicity we explicitly report only the more important proton case)
\begin{equation}
\sigma_{\chi_1 p}^{\rm SI}=\frac{\mu_{\chi_1 p}^2}{\pi M_H^4}|g_{H \chi_1 \chi_1}|^2\frac{m_p^2}{v^2} {\left[f_p {Z}/{A}+f_n \left(1-{Z/}{A}\right)\right]}^2,
\end{equation}
$\mu_{\chi_1 p} =m_{\chi_1} m_p/(m_{\chi_1}+ m_p)$  is the DM/proton reduced mass. $f_p \simeq f_n \approx 0.3$ are the effective couplings of the DM with the nucleons. $A,Z$ represent the atomic number and the number of protons of the element/material composing a given detector; at the moment the reference constraints are provided by Xenon based experiments such as LZ and XENON. It is useful to report the explicit expression of the DM-Higgs coupling
\begin{equation} 
g_{H \chi_1 \chi_1}=-\frac{y^2 v \left(m_{\chi_1} +m_D \sin 2\theta\right)}{m_D^2 + 2 m_D m_{\chi_1}-3 m_{\chi_1}^2  +y^2 v^2/2} \, , 
\end{equation}
from which one can see that it can be set to zero if the term $m_{\chi_1}+m_D \sin2\theta$ vanishes. If it is indeed the case, a so-called blind spot \cite{Choudhury:2015lha,Choudhury:2017lxb} occurs for these spin-independent interactions. Spin-dependent interactions are instead due to the DM axial-vector interactions with the $Z$ boson. The corresponding cross section is given by
\begin{equation}
\sigma_{N_1 p}^{\rm SD}=\frac{\mu_{\chi_1 p}^2}{\pi M_Z^4}|g_{Z \chi_1 \chi_1}^A|^2 {\left[A_u^{Z} \Delta_u^p+ A_d^Z \left(\Delta_d^p+\Delta_s^p\right)\right]}^2 \, ,
\end{equation}
A blind spot $g_{Z \chi_1 \chi_1}^A\!=\!0$ can also occur for spin-dependent interactions when $|U_{12}|\!=\!|U_{13}|$. Even if the singlet-doublet model is potentially also testable in indirect detection, as some of the relevant annihilation processes like those into $W/Z$ bosons are $s$-wave dominated, the corresponding limits are not competitive with the ones from direct detection. They will thus not be explicitly reported here and for more details, see eventually Refs.~\cite{Calibbi:2015nha,Arcadi:2017kky}. 

\subsection{The CDF W-mass anomaly and the new fermionic sector}

We come now to the discussion of the new contributions to the the $W$ boson mass and confront them with the recent CDF measurement. At leading order, the variation of the electroweak observables and in particular $M_W$ with respect to the SM prediction, can be related to a deviation from the custodial limit $\Delta \rho= 1/(\rho-1)=0$ of the $\rho$ parameter which measures the strength of the neutral to charged currents ratio at zero-momentum transfer \cite{Veltman:1977kh,Toussaint:1978zm}: ${\Delta M_W}/{M_W} \approx \frac{3}{4} \Delta \rho$. The contribution to $\Delta \rho$ (and hence to $\Delta M_W$ and other observables) of two particles of an SU(2) doublet with masses that have a large splitting can be rather large as it is quadratic in the mass of the heaviest particle \cite{Veltman:1977kh}. 

To take also into account subleading contributions to $\Delta M_W$, one can e.g. consider the Peskin-Takeuchi approach with the $S,T,U$ parameters \cite{Peskin:1991sw}. In this scheme, the largest contribution $T$ is in fact simply $\Delta \rho$, $T \propto \Delta \rho -\Delta \rho|_{\rm  SM}$,   while $S$ describes new contributions from neutral current processes at different energies and $U$ the contribution to $M_W$ from new charged currents (this last correction is in general small and we will neglect it here). 
In our singlet-doublet model, the contributions to the $S$ and $T$ parameters originate from the new fermionic sector that couples to the $W$ and $Z$ bosons \cite{He:2001tp,Barbieri:2006bg,Enberg:2007rp,DEramo:2007anh,Joglekar:2012vc}. 

The new fermion (NF) contributions  can be schematically written as \cite{DEramo:2007anh}
\begin{eqnarray}
   \Delta S_{\rm NF}&=&\Sigma_{i,j=1}^3 {\left(U_{1i}U_{2j}+U_{2i}U_{1j}\right)}^2 {F}(m_{\chi_i},-m_{\chi_j})- {F}(m_D,m_D) \, , \nonumber\\
   \Delta T_{\rm NF}&=&\Sigma_{i=1}^3 \left[(U_{1i})^2 {G}(m_S,m_{\chi_i})+(U_{2i})^2 {G}(m_S,-m_{\chi_i})\right] \, , 
\end{eqnarray}
where the functions $F$ and $G$ are given by ($\alpha_{\rm EM}$ is the fine structure constant)  
\begin{eqnarray}
F (m_A,m_B)= & \frac{1}{6\pi (m_A^2-m_B^2)^2}  \bigg[ m_A m_B (3 m_A^2-4 m_A m_B+3 m_B^2 + \frac{1}{m_A-m_B} \times \nonumber\\     
    & [m_A^6+m_B^6-3 m_A^2 m_B^2 (m_A^2+m_B^2)+6 m_A^3 m_B^3] \bigg] , \\
     {G}(m_A,m_B)=& \frac{1}{16\pi^2 \alpha_{\rm EM}v^2}\left[-2 m_A m_B +\frac{2 m_A m_B (m_A^2+m_B^2)-m_A^4-m_B^4}{m_A^2-m_B^2}\log\frac{m_A^2}{m_B^2}\right] \, . \nonumber
\end{eqnarray}

\begin{figure}[!ht] 
\vspace*{-1mm} 
    \centering
  \includegraphics[width=0.45\linewidth]{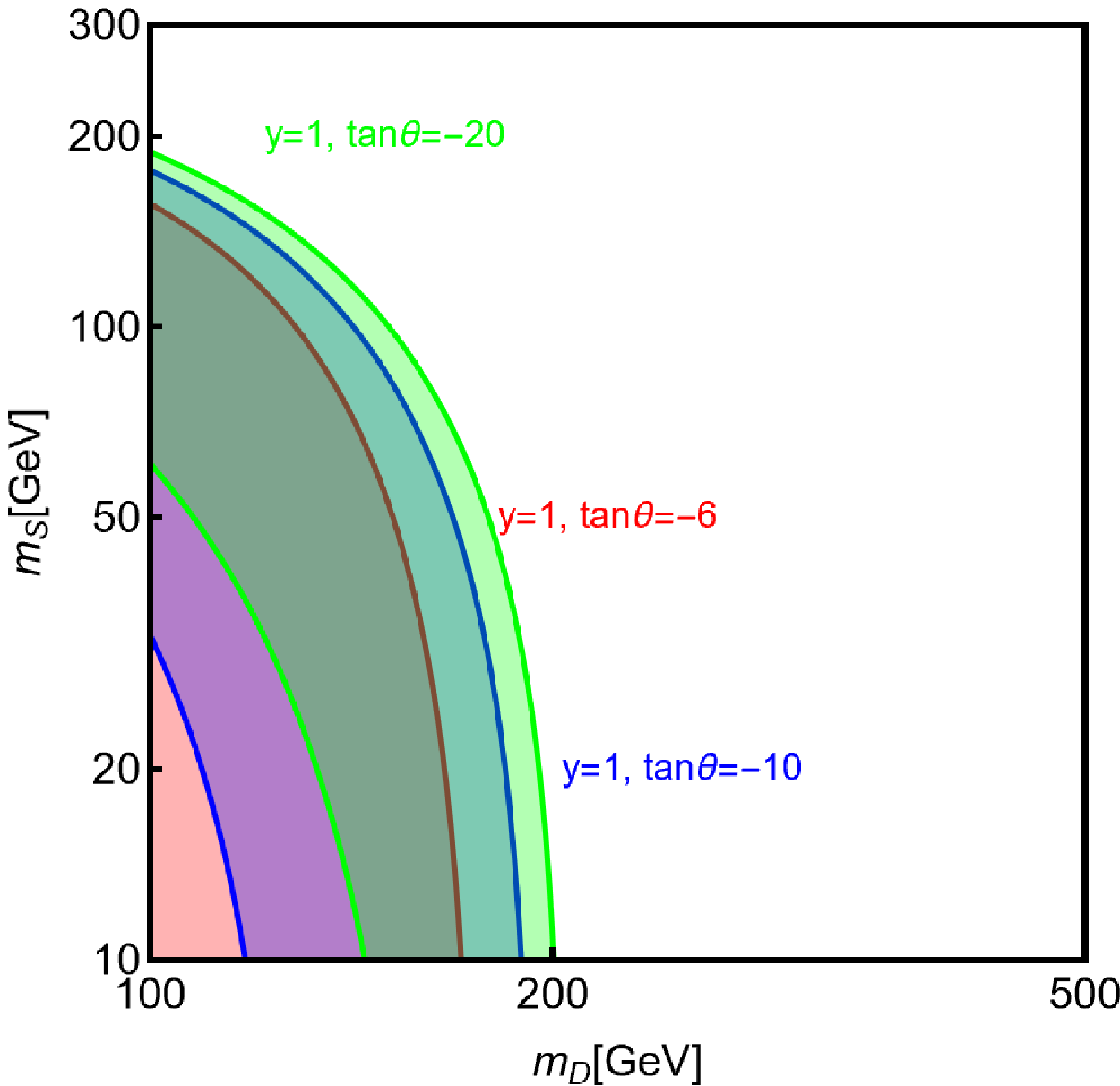} ~~
  \includegraphics[width=0.44\linewidth]{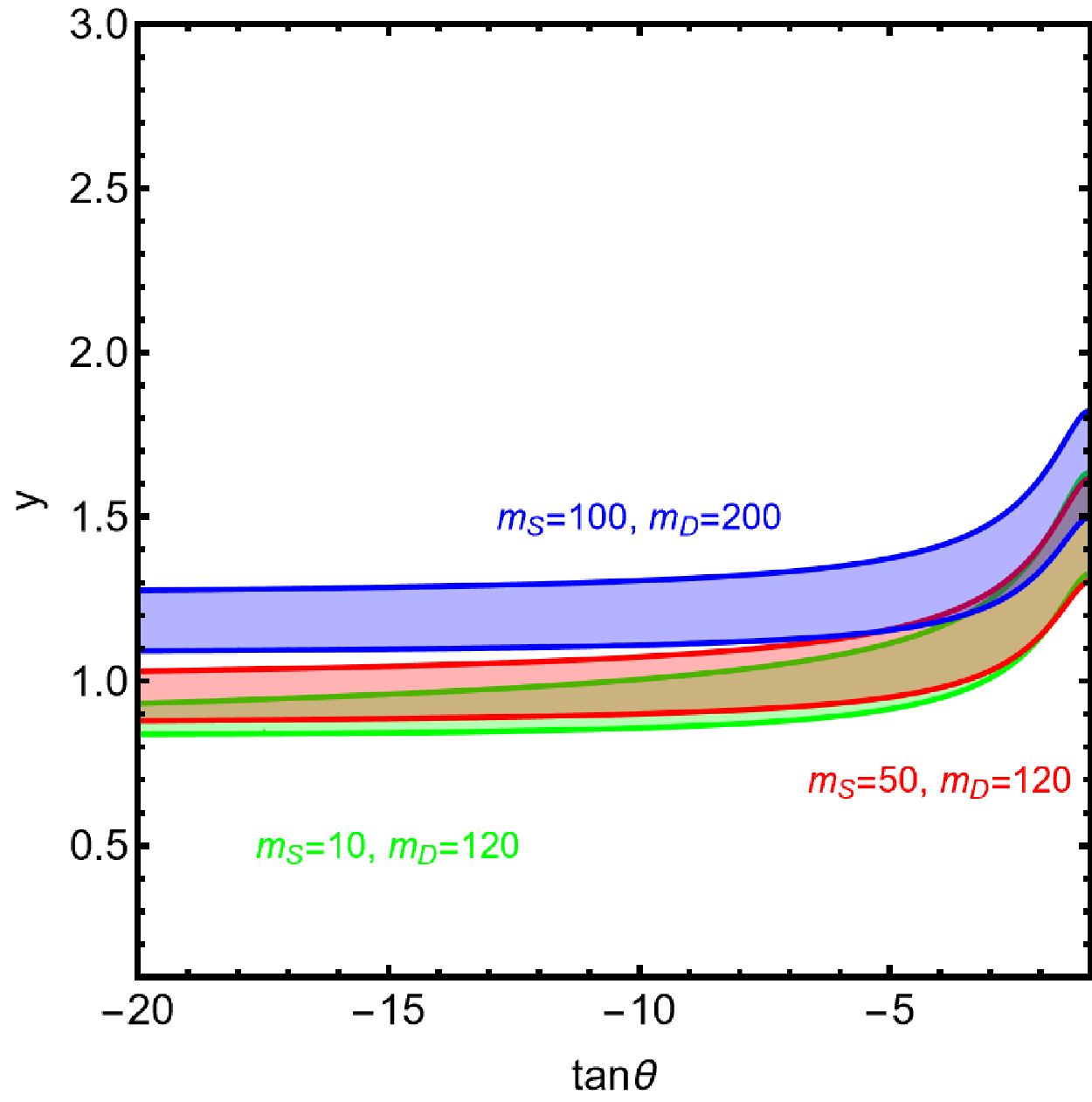}  
     \vspace*{-2mm}  
    \caption{Regions of the $[m_D,m_S]$ (left) and $[\tan\theta, y]$ (right) planes for the singlet-doublet model with a SM-like Higgs sector which comply with the $M_W^{\rm CDF}$ anomaly. The different colors correspond to the choices of $(y,\tan\theta)$ or $(m_S, m_D)$ given in the plots.}
    \label{fig:pmwSD}
     \vspace*{-2mm}  
\end{figure}

As an illustration,  we show in Fig.~\ref{fig:pmwSD} the regions of the $[m_D,m_S]$ and $[y,\tan\theta]$ planes which provide a viable fit of the $M_W^{\rm CDF} $ anomaly. The three different colors of the contours correspond to  the three assignments of the $(y,\tan\theta)$ pairs, namely $(1,-6)$, $(1,-10)$ and $(1,-20)$ in the left plot and $(m_S,m_D)$ pair, namely  $(10, 120), (50, 100)$ and $100,200)$ [in GeV] on the right plot. We restricted to mass values $m_D \geq 100$ GeV to comply with limits on charged leptons from the LEP experiment \cite{ParticleDataGroup:2020ssz}.  The reason for the negative values of $\tan\theta$ is that they allow for a blind spot in DM direct detection as will be seen later.

As evidenced by Fig. \ref{fig:pmwSD}, the CDF $M_W$ measurement seems to favor relatively low values of the masses $m_S$ and $m_D$, implying a rather light DM candidate. In such a case, a relevant complementary constraint would be represented by the invisible widths of the $Z$ and $H$ bosons, as both particles can decay into a pair of the escaping DM candidate, if such processes are kinematically allowed, i.e when  $m_{\chi_1} <\frac12 M_Z$ and 
$m_{\chi_1} <\frac12 M_H$ respectively. 

Additional decay processes of the $Z$ boson are  strongly constrained by precision measurements performed at LEP, which can be summarized by the upper bound $\Gamma(Z \rightarrow \mbox{inv})< 2.3 \,\mbox{MeV}$ for $m_S \leq 45$ GeV \cite{ParticleDataGroup:2020ssz}. 
Likewise, extra exotic decays of the 125 GeV Higgs boson are disfavored by LHC measurements of the $H$ couplings to fermions and gauge bosons.  The most recent  results lead to an upper bound on the Higgs invisible decay branching ratio  of BR$(H \rightarrow \,\mbox{inv})< 0.11$ \cite{ATLAS-new,CMS-new}. Additional constraints on the masses $m_S$ and $m_D$, and on the parameters $y$ and $\theta$,   could come from direct searches at LEP2 and at the LHC but they are model dependent and we will ignore them here for simplicity. 

Before moving to the combination of our results, we note that as the new fermionic sector does not couple or mix with to SM fermions, it does not contribute to $(g-2)_\mu$ and, thus,  the anomalous Fermilab result cannot be explained in this minimal model.

\subsection{Combined numerical results}

\begin{figure}[!ht] 
\vspace*{-2mm}
    \centering
\mbox{\includegraphics[width=0.48\linewidth]{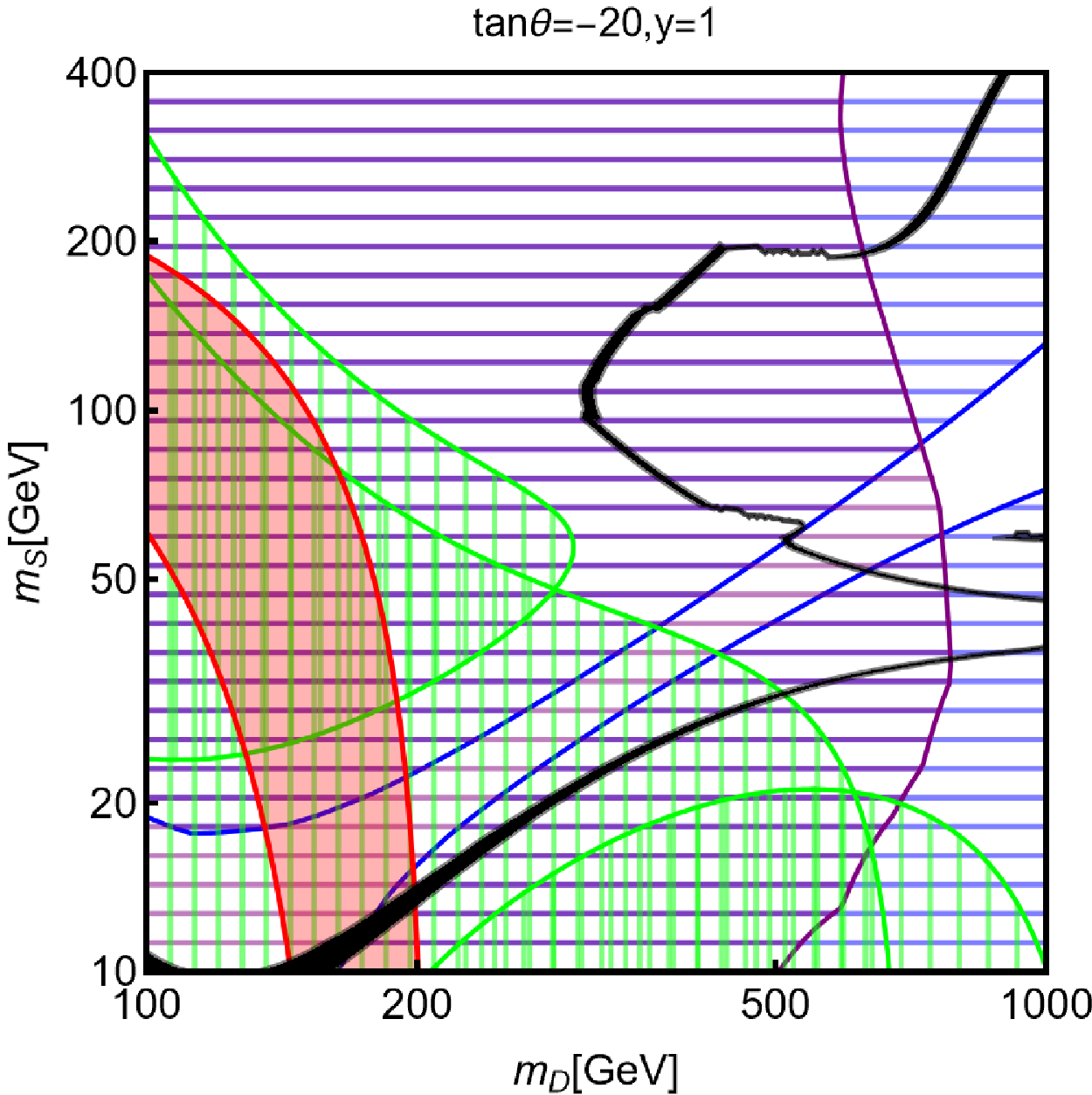}~~
      \includegraphics[width=0.48\linewidth]{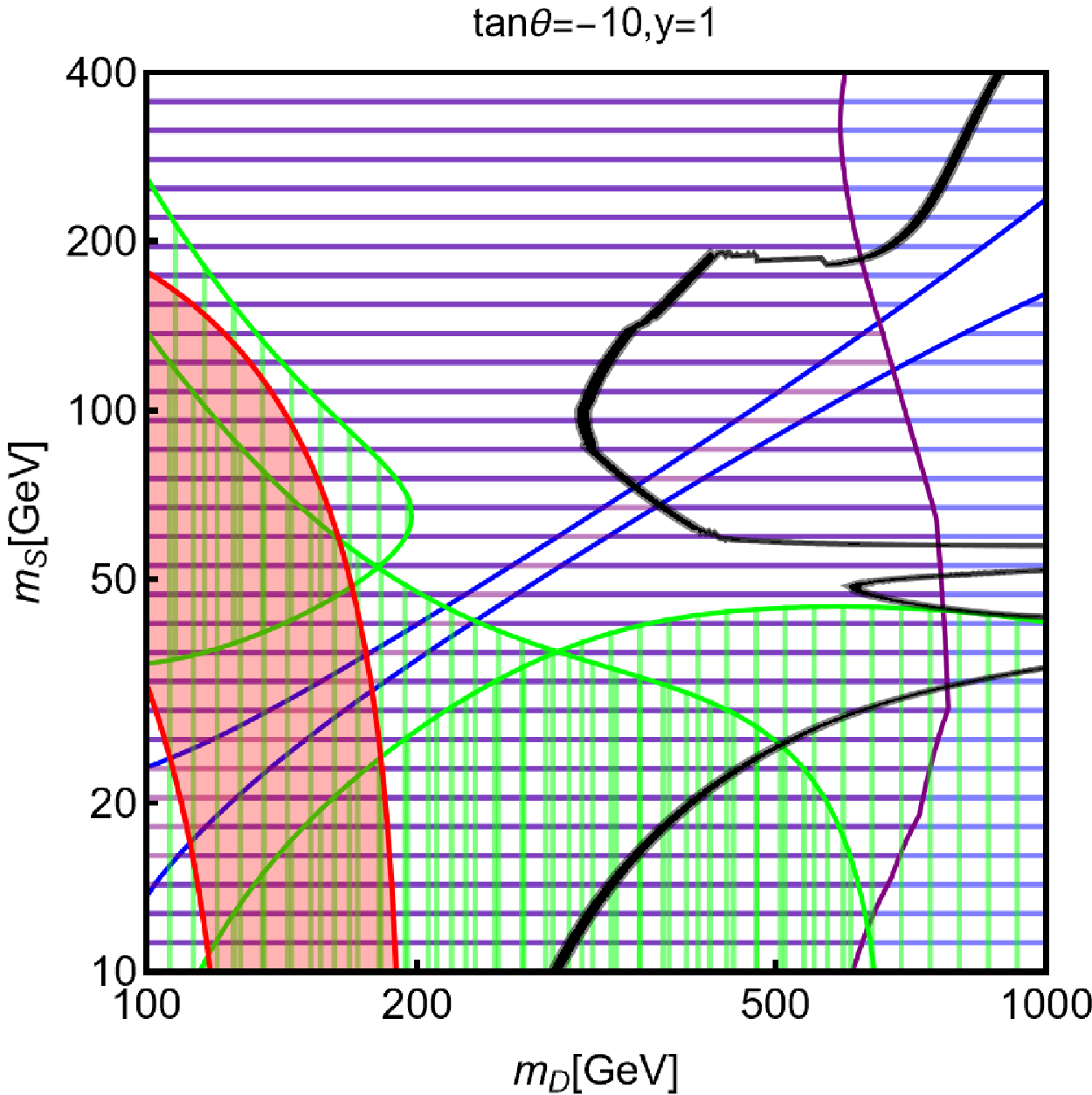} }
\vspace*{-2mm}
    \caption{Summary of constraints for the singlet-doublet model: the black isocontour corresponds to the correct relic density while the red regions provide a viable interpretation of the $M_W^{\rm CDF}$ discrepancy. The hatched regions correspond to different experimental exclusions, namely limits from spin-independent interactions (blue), spin-dependent interactions (purple) and the invisible widths of the $H/Z$ bosons (green).}
    \label{fig:plotSD}
\vspace*{-2mm}
\end{figure}

We have now all the elements to discuss our main numerical results that combine all collider and astroparticle physics constraints, which are  reported in Fig. \ref{fig:plotSD}. 
The figure compares the regions of parameter space in the $[m_D,m_S]$ plane accounting for the $M_W^{\rm CDF}$ anomaly and including the different constraints. More precisely, the black isocontours represent the viable relic density according to the standard WIMP paradigm, while the hatched regions correspond to the various experimental exclusion bounds.

The blue and purple regions are excluded, respectively, by limits from spin-independent and spin-dependent DM interactions according to the most recent determination made by the LZ \cite{LZ:2022ufs} and XENON1T \cite{XENON:2019rxp} experiments. The green regions are, instead, excluded by searches of invisible decay branching fractions of the SM Higgs and $Z$ bosons. A given benchmark will be regarded as viable if there is a non-zero intersecting area between the red and black contours and outside the colored regions corresponding to the  experimental exclusions. As can be seen, this is not the case of the benchmarks shown in the figure. This is mostly due to the very strong constraints from DM detection which rule out most of the $[m_D,m_S]$ plane. Indeed, given the different interactions responsible for the spin-independent and spin-dependent cross sections, it is very difficult to achieve blind spots for both of them at the same time. Given also the low values of the masses $m_{S}$ and $m_{D}$ that are needed to reproduce the CDF $M_W^{}$ anomaly, at least one of the two limits (together with the ones from invisible $H/Z$ decays) becomes effective.\vspace*{-1mm} 

An extension of the model is thus required to bypass these limitations. An  interesting possibility would be a Higgs sector with two doublets  to which we turn our attention now. 

%%%%%%%%%%%%%%%%%%%%%%%%%%%%%%%%%%%%%%%%%%%%%%%%%%%%%%%%%%%

\section{The singlet-doublet fermion case in a 2HDM}

\subsection{The 2HDM and its ingredients} 
 
We consider the case that the scalar sector of the theory is composed by two doublet fields $\Phi_1$ and $\Phi_2$ which lead to the CP-conserving $Z_2$ invariant potential~\cite{Davidson:2005cw}
\begin{align}
\label{eq:scalar_potential}
 V(\Phi_1,\Phi_2) &\!= \!m_{11}^2 \Phi_1^\dagger \Phi_1\!+\! m_{22}^2 \Phi_2^\dagger \Phi_2 \!-\! m_{12}^2 \left(\Phi_1^\dagger \Phi_2 \!+\! {\rm h.c.} \right) \!+\! \frac{\lambda_1}{2} \left( \Phi_1^\dagger \Phi_1 \right)^2
\!+\!\frac{\lambda_2}{2} \left( \Phi_2^\dagger \Phi_2 \right)^2 \hspace*{-1cm}  \nonumber\\ &
\!+\!\lambda_3\left(\Phi_1^\dagger \Phi_1 \right)\left(\Phi_2^\dagger \Phi_2 \right)
\!+\!\lambda_4\left(\Phi_1^\dagger \Phi_2 \right)\left(\Phi_2^\dagger \Phi_1 \right)
\!+\!\frac{\lambda_5}{2}\left[ \left(\Phi_1^\dagger \Phi_2 \right)^2 \!+\! {\rm h.c.} \right] .
\end{align}
After electroweak symmetry breaking, the two doublets can be decomposed as
\begin{equation}
\Phi_i=
\begin{pmatrix} \phi_i^+ \\ (v_i+\rho_i +i \eta_i)/\sqrt{2} \end{pmatrix}~,
\qquad
i=1,2,
\end{equation}
\noindent
with $v_{1}$ and $v_2$ the vacuum expectation values satisfying $\sqrt{v_1^2\!+\! v_2^2}\!=\!v \simeq 246$ and $\tan\beta \! = \! v_2/v_1$.  The physical mass eigenstates emerge via rotations with angles $\alpha$ and $\beta$  
\begin{equation}
\label{eq:rotation2}
\left(
\begin{array}{c} \phi_1^+ \\ \phi_2^+ \end{array} \right) = \Re_\beta 
\left( \begin{array}{c} G^+ \\ H^+ \end{array} \right), \ \  
%\nonumber\\
\left( \begin{array}{c} \eta_1 \\ \eta_2 \end{array} \right)= \Re_\beta 
\left( \begin{array}{c} G^0 \\ A \end{array} \right), \ \
\left( \begin{array}{c} \rho_1 \\ \rho_2 \end{array} \right)= \Re_\alpha 
\left( \begin{array}{c} H \\ h \end{array} \right) \, , 
\end{equation} 
with $\Re_{X,X=\alpha,\beta}$ being the rotation matrices of angle $X$ with elements given in terms of $\cos X$ and $\sin X$. The states $h,H$ are the neutral CP-even Higgs bosons with $h$ being identified with the observed 125 GeV Higgs state; throughout this work we will assume the hierarchy $M_h < M_H$. $A$ is a CP-odd Higgs eigenstate while $H^{\pm}$ are the electrically charged Higgs states. Finally, $G^0$ and $G^+$ are the Goldstone bosons that make the longitudinal degrees of freedom of the $Z$ and $W$ bosons.

In order to be theoretically consistent, the parameters of the scalar potential should comply with a series of constraints (see for instance Ref.~\cite{Barroso:2013awa}): $i)$ the scalar potential should be bounded from below, $ii)$ it obeys $s$-wave unitarity at the tree level, $iii)$ the electroweak vacuum should be a stable global minimum, and finally $iv)$ the couplings should stay perturbative, i.e. $|\lambda_i|\leq 4 \pi$. These constraints have been discussed in the literature and a recent account has been given e.g. in Ref.~\cite{Arcadi:2022lpp}. They can be translated into constraints on the masses of the various Higgs mass eigenstates using relations also given in Ref.~\cite{Arcadi:2022lpp}. We will include all these constraints in our numerical analysis. 

Turning to the couplings between the physical Higgs bosons and the SM fermions, they are described by the following Yukawa-type Lagrangian
\begin{align}
-{\mathcal L}_{\rm Yuk}^{\rm SM}& =\Sigma_{f=u,d,l} \frac{m_f}{v} \left[g_{hff} \bar{f}f h +g_{Hff} \bar{f}f H-i g_{Aff} \bar{f} \gamma_5 f A \right] \notag \\
&- ({\sqrt{2}}/{v}) \left[ \bar{u} \left(m_u g_{Auu} P_L + m_d g_{Add} P_R \right)d H^+ +  m_l g_{All} \bar \nu  P_R \ell H^+  + \mathrm{h.c.} \right],
\end{align}

with $P_{L/R}\!=\!\frac12(1\!\mp\!\gamma_5)$ and  $g_{\phi ff}$ the reduced couplings of the $\phi$ boson to up- and down-type quarks and charged leptons normalized to the SM couplings, $g_{\phi ff}\!=\!g^\text{2HDM}_{\phi ff}/g^\text{SM}_{H ff}$. 

\begin{table}[h!]
\vspace*{-3mm}
\renewcommand{\arraystretch}{1.2}
\begin{center}
\begin{tabular}{|c|c|c|c|c|}
\hline
~~~~~~ &  ~~Type-I~~ & ~~Type-II~ & ~~Type-X~~ & ~~Type-Y~~ \\
\hline 
$g_{huu}$ & $ \frac{\cos \alpha} { \sin \beta} \rightarrow 1$ & $\frac{ \cos \alpha} {\sin \beta} \rightarrow 1$ & $\frac{ \cos \alpha} {\sin\beta} \rightarrow 1$ & $ \frac{ \cos \alpha}{ \sin\beta}\rightarrow 1$ \\ \hline
$g_{hdd}$ & $\frac{\cos \alpha} {\sin \beta} \rightarrow 1$ & $-\frac{ \sin \alpha} {\cos \beta} \rightarrow 1$ & $\frac{\cos \alpha}{ \sin \beta} \rightarrow 1$ & $-\frac{ \sin \alpha}{ \cos \beta} \rightarrow 1$ \\ \hline
$g_{h\ell \ell} $ & $\frac{\cos \alpha} {\sin \beta} \rightarrow 1$ & $-\frac{\sin \alpha} {\cos \beta} \rightarrow 1$ & $- \frac{ \sin \alpha} {\cos \beta} \rightarrow 1$ & $\frac{ \cos \alpha} {\sin \beta} \rightarrow 1$  \\ \hline
$g_{Huu}$ & $\frac{\sin \alpha} {\sin \beta} \rightarrow -\frac{1}{\tan\beta}$ & $\frac{ \sin \alpha} {\sin \beta} \rightarrow -\frac{1}{\tan\beta}$ & $ \frac{\sin \alpha}{\sin \beta} \rightarrow -\frac{1}{\tan\beta}$ & $\frac{ \sin \alpha}{ \sin \beta} \rightarrow -\frac{1}{\tan\beta}$ \\ \hline
$g_{Hdd}$ & $ \frac{ \sin \alpha}{\sin \beta} \rightarrow -\frac{1}{\tan\beta}$ & $\frac{\cos \alpha}{\cos \beta} \rightarrow {\tan\beta}$ & $\frac{\sin \alpha} {\sin \beta} \rightarrow -\frac{1}{\tan\beta}$ & $\frac{ \cos \alpha} {\cos \beta} \rightarrow {\tan\beta}$ \\ \hline
$g_{H\ell\ell}$ & $\frac{ \sin \alpha} {\sin \beta} \rightarrow -\frac{1}{\tan\beta}$ & $\frac{\cos \alpha} {\cos \beta} \rightarrow {\tan\beta}$ & $\frac{ \cos \alpha} {\cos \beta} \rightarrow {\tan\beta}$ & $\frac{\sin \alpha} {\sin \beta} \rightarrow -\frac{1}{\tan\beta}$ \\ \hline
$g_{Auu}$ & $\frac{1}{\tan\beta}$ & $\frac{1}{\tan\beta}$ & $\frac{1}{\tan\beta}$ & $\frac{1}{\tan\beta}$ \\ \hline
$g_{Add}$ & $-\frac{1}{\tan\beta}$ & ${\tan\beta}$ & $-\frac{1}{\tan\beta}$ & ${\tan\beta}$ \\ \hline
$g_{A\ell \ell}$ & $-\frac{1}{\tan\beta}$ & ${\tan\beta}$ & ${\tan\beta}$ & $-\frac{1}{\tan\beta}$
\\
\hline
\end{tabular}
\caption{Couplings of the 2HDM Higgs bosons to fermions, normalized to those of the SM-like Higgs boson,  as a function of the angles $\alpha$ and $\beta$. In the case of the CP-even Higgs states,  their values in the alignment limit $\beta \!-\! \alpha \rightarrow \frac{\pi}{2}$.}
\label{table:2hdm_type}
\end{center}
\vspace*{-6mm}
\end{table}

To avoid the emergence of tree-level flavour-changing neutral currents, only four possible sets of assignments of the couplings can be considered \cite{Branco:2011iw,Glashow:1976nt}; they are dubbed Type-I, Type-II, Type-X (or lepton specific) and Type-Y (or flipped) 2HDMs. The corresponding couplings are summarized in Table \ref{table:2hdm_type}. Note that the angle $\alpha$, which determines the mixing between the neutral CP-even states $h$ and $H$, is constrained by the measurement of the couplings of the $h$ state at the LHC which should be SM-like. The statement is enforced quantitatively by allowing only small deviations from the so-called alignment limit $\beta-\alpha=\frac{\pi}{2}$, see e.g. Ref.~\cite{Pich:2009sp}. As for the couplings of the other Higgs states, it can be seen from the table that they can be strongly enhanced or suppressed with respect to the SM values, depending on the value of $\tan\beta$ and the considered configuration. Consequently, different experimental limits should apply in the different cases and we refer to e.g. Ref.~\cite{Arcadi:2019lka} for a review. As discussed in Ref.~\cite{Arcadi:2021zdk,Arcadi:2022dmt}, in order to comply with the $(g-2)_\mu$ and $M_W^{\rm CDF}$ anomalies, one should focus on the lepton specific or Type-X 2HDM, with large values of the parameter $\tan\beta$ to enhance the lepton couplings. It allows to achieve a relatively light spectrum for the additional Higgs states, while still complying with most of the bounds coming from collider searches and flavour physics.

Considering finally the interactions of the Higgs sector with the singlet-doublet fermionic states, the relevant Lagrangian is a straightforward generalization of the one presented in the previous section and can be written as   ($a,b=1,2$) \cite{Berlin:2015wwa,Arcadi:2018pfo}
\begin{eqnarray}
\mathcal{L}=-\frac12m_S S^{2}-m_D D_L D_R -y_1 D_L \Phi_a S-y_2 D_R \widetilde{\Phi}_b S^+\mbox{h.c.}
\end{eqnarray}
The fermionic physical eigenstates will be still represented by three neutral Majoranas and one electrically charged Dirac fermion. This time, the neutral mixing matrix will depend on the two different vevs $v_1$ and $v_2$.  Consequently, the singlet and doublet components of the DM as well as its couplings will be sensitive also to the angles $\alpha$ and $\beta$, in addition to the masses $m_S$ and $m_D$. In the fermion mass basis, the interaction Lagrangian reads
\begin{eqnarray}
 \mathcal{L} \! &\!=\!\bar{\psi^-} \gamma^\mu \left(g^V_{W^{\mp}\psi^{\pm} \chi_i}\!-\!g^A_{W^{\mp}\psi^{\pm}N_i}\gamma_5\right)\chi_i W_\mu^{-}
\!+\!\frac{1}{2}\Sigma_{i,j=1}^3 \bar{\chi_i}\gamma^\mu \left(g_{Z \chi_i \chi_j}^V\!-\!g_{Z \chi_i \chi_j}^A \gamma_5\right) \chi_j Z_\mu \hspace*{-1cm} \nonumber\\
& \!+\!\frac12 \Sigma_{i,j=1}^{3}\bar{\chi_i}\left(y_{h \chi_i \chi_j}h\!+\! y_{H \chi_i \chi_j}H\!+\!y_{A \chi_i \chi_j}\gamma_5 A\right)\chi_j \!+\!\bar{\psi^-} \left(g^S_{H^{\pm}\psi \chi_i}\!-\!g^P_{H^{\pm}\psi \chi_i}\gamma_5\right)\chi_i H^{-} \hspace*{-1cm} \nonumber\\
& -e A_\mu \bar{\psi^{-}}\gamma^\mu \psi^{-}-\frac{g}{2 \cos\theta_W}(1-2 \sin^2\theta_W) Z_\mu \bar{\psi^{-}}\gamma^\mu \psi^{-}+\mbox{h.c.} , 
\end{eqnarray}
\noindent
where the Higgs couplings in the case of $\phi=h,H,A$ and $H^\pm$ are given by
\begin{align}
\label{eq:SD2HDM_couplings}
& y_{ \phi \chi_i \chi_j}=\frac{\delta_\phi}{2\sqrt{2}}\left[U_{i1}\left(y_1 R_a^\phi U_{i2}+y_2 R_b^\phi U_{i3}\right)+(i \leftrightarrow j)\right] ,  \nonumber\\
& g^{S/P}_{H^{\pm}\psi \chi_i}=\frac{1}{2}U_{i1}\left(y_1 R_1^{H^{\pm}} \pm y_2 R_2^ {H^{\pm}}\right) ,
\end{align}
with $\delta_h=\delta_H=-1$ and $\delta_A=-i$. Similarly to what occurs for the SM fermions, one should not assume arbitrary couplings of the new fermions with the  $\Phi_1$ and $\Phi_2$ doublet fields. The simplest way to proceed would consist of extending to the new fermionic sector the same symmetries which define the four flavor conserving 2HDMs defined earlier \cite{Berlin:2015wwa,Arcadi:2018pfo}. This leads to two possible assignments of the $R_{a,b}^\phi$ parameters 
\begin{eqnarray}
\label{eq:assA}
 R_1^h\!=\!R_2^h\!=\!-\!\sin\alpha, \ R_1^H\!=\!R_2^H\!=\!\cos\alpha, \ R_1^A\!=\! R_2^A\!=\!-\! \sin\beta, \ R_1^{H^{\pm}}\!=\!R_2^{H^{\pm}}\!=\!-\!\sin\beta \\
\label{eq:assB}
R_1^h\!=\!-\!R_2^H \!=\! -\!\sin\alpha, \ R_2^h\!=\!R_1^H\!=\! \cos\alpha,\  R_1^A\!=\!R_1^{H^{\pm}}\!=\!-\!\sin\beta,\ R_2^A\!=\!R_2^{H^{\pm}} \!= \!\cos\beta . 
\end{eqnarray}
which will be dubbed Type-A for the first one and Type-B for the second configuration.

 This completes the necessary ingredients to study the phenomenology of the model.

\subsection{The DM sector}

The phenomenology of the DM particle in the present case bears many similarities with the already discussed minimal singlet-doublet model. We thus simply point out the additional features that are due to the extended Higgs sector. Starting with DM direct detection, the spin-independent cross section receives an additional contribution from the $t$-channel exchange of the heavy CP-even $H$ state, and will be then given by \cite{Berlin:2015wwa,Arcadi:2018pfo}
\begin{equation}
\sigma_{\chi p}^{\rm SI}=\frac{\mu_{\chi\,p}^2}{\pi}\frac{m_p^2}{v^2}\bigg| \sum_{q}f_q \left(\frac{y_{h \chi_1 \chi_1}g_{hqq}}{M_h^2}+\frac{y_{H \chi_1 \chi_1}g_{Hqq}}{M_H^2}\right) \bigg|^2 \ .
\end{equation}
In contrast, the functional form of the spin-dependent cross section is unchanged with respect to the one in the minimal model presented in section 2.2. Again, the spin-inde\-pen\-dent cross section can be set to zero at tree-level by choosing vanishing Higgs couplings $y_{h \chi_1 \chi_1}\!=\!y_{H \chi_1 \chi_1}\!=\!0$ by imposing the relation $m_S\!+\!m_D \sin 2 \theta \! \simeq \! 0$. A blind-spot can also be generated with a destructive interference between the $h/H$-exchange contributions.

For what concerns the relic density, there are mostly two relevant changes with respect to the minimal model. First, we have the possibility of the extra $s$-channel exchange of the pseudoscalar boson $A$ (in addition to that of the $H$ state) in DM annihilation into SM fermions final states. This additional contribution has no counterpart in the interactions relevant for DM direct detection and, hence, could potentially alleviate the tensions that are present in the minimal model. A further relevant impact on the DM relic density would appear when one of the extra Higgs bosons is lighter than the DM particle,  implying  the possibility of additional annihilation channels for the latter.

Finally, there are also bounds on the DM mass and couplings from collider searches as already discussed in the previous section when we considered 
the invisible decay widths of the $h$ and $Z$ bosons that would also apply in  the 2HDM realization. One additional feature not present earlier is that, in the case of a light pseudoscalar state, the width of the 125 GeV Higgs boson can get additional exotic contributions corresponding to the $h \!\to \! ZA$ and $h\to \! AA$ channels. The former is absent in the alignment limit as $g_{hZA}=0$ and the latter is subject to a very active search program at the LHC, see e.g. Refs.~\cite{ATLAS:2020ahi,CMS:2020ffa,CMS:2021pcy}
%ATLAS:2021shl,CMS:2022fyt}. 
The partial decay width of the 125 Higgs into two light pseudoscalars is given by \cite{Djouadi:2005gj}
\begin{equation}
    \Gamma (h\! \rightarrow \! AA)\! =\! \frac{\left \vert \lambda_{hAA}\right \vert^2}{32\pi M_h}\sqrt{1-{4 M_A^2}/{M_h^2}}\, , 
\end{equation}
where, using the abbreviation $M^2 \equiv m_{12}^2/(\sin {\beta} \cos {\beta})$ with $m_{12}$ appearing in the 2HDM scalar potential given in eq.~(\ref{eq:scalar_potential}), one has 
\begin{equation}
  \hspace*{-5mm}   \lambda_{hAA}\!=\! \frac{1}{2v}\left[\left(2 M^2\!-\! 2 M_A^2\!-\!M_h^2\right) \sin(\beta \!-\! \alpha)\!+\! \left(M^2\!-\!M_h^2\right)(\cot \beta\!-\! \tan\beta)\cos(\beta \!-\!\alpha)\right]\! .~~ \hspace*{-1cm}
\end{equation}
One can see that it is possible to set the $\lambda_{hAA}$ coupling to zero, i.e. to achieve a kind of blind spot, by imposing the relation \cite{Abe:2015oca}
\begin{equation}
\label{eq:blindspot}
    \tan(\beta-\alpha)=\frac{M^2-M_h^2}{2 M^2-2 M_A^2-M_h^2}(\tan\beta-\cot \beta) . 
\end{equation}

\subsection{Interpreting the CDF W-mass anomaly}

We come now to the contributions of the new particles of this extended singlet-doublet scenario to the electroweak observables and, in particular, to the mass $M_W$.  Besides the new fermion contributions to the $S$ and $T$ parameters, which have exactly the same functional form given in section 2.3, one needs to include those of the extended Higgs sector. The contribution to the $S$ and $T$ parameters from a 2HDM can be written as \cite{Barbieri:2006bg}
\begin{eqnarray}
    S_{\rm 2HDM}&\!=\!& F(M_h,M_Z)+M_Z^2 G(M_h,M_Z) +F(M_A,M_H) 
    -F(M_{H^{\pm}},M_{H^{\pm}}) \, ,  \nonumber \\ 
     T_{\rm 2HDM}&\!=\!& \!-\!3  [A(M_h,M_W)\!-\!A(M_h,M_Z) ]\!+\!      
     F(M_A,M_H) \!-\!F(M_{H^{\pm}},M_{H^{\pm}}) \, , 
\end{eqnarray}
where we have assumed the alignment limit $\alpha=\beta- \frac{\pi}{2}$ and used the functions
\begin{eqnarray}
     A(m_A,m_B)&=& \frac{1}{32\pi^2 \alpha_{\rm EM}v^2}\left[\frac{m_A^2+m_B^2}{2}-\frac{m_A^2 m_B^2}{m_A^2-m_B^2}\log \frac{m_A^2}{m_B^2}\right] , \nonumber\\
    F(m_A,m_B)&=&\frac{1}{24\pi}\left[ \frac{4 m_A^2 m_B^2}{( m_A^2-m_B^2)^2} +\frac{m_A^6+m_B^6-3 m_A^2 m_B^2 (m_A^2+m_B^2)}{(m_A^2-m_B^2)^3} \log \frac{m_A^2}{m_B^2} \right] \, , \nonumber\\
    G(m_A,m_B)&=&\frac{1}{2\pi}\left[\frac{2m_A^2 m_B^2}{{\left(m_A^2-m_B^2\right)}^3}\log \frac{m_A^2}{m_B^2}-\frac{m_A^2+m_B^2}{{\left(m_A^2-m_B^2\right)}^2}\right]\, . 
\end{eqnarray}

\begin{figure}[!ht]  
 \vspace*{-1mm}
    \centering
    \includegraphics[width=.43\linewidth]{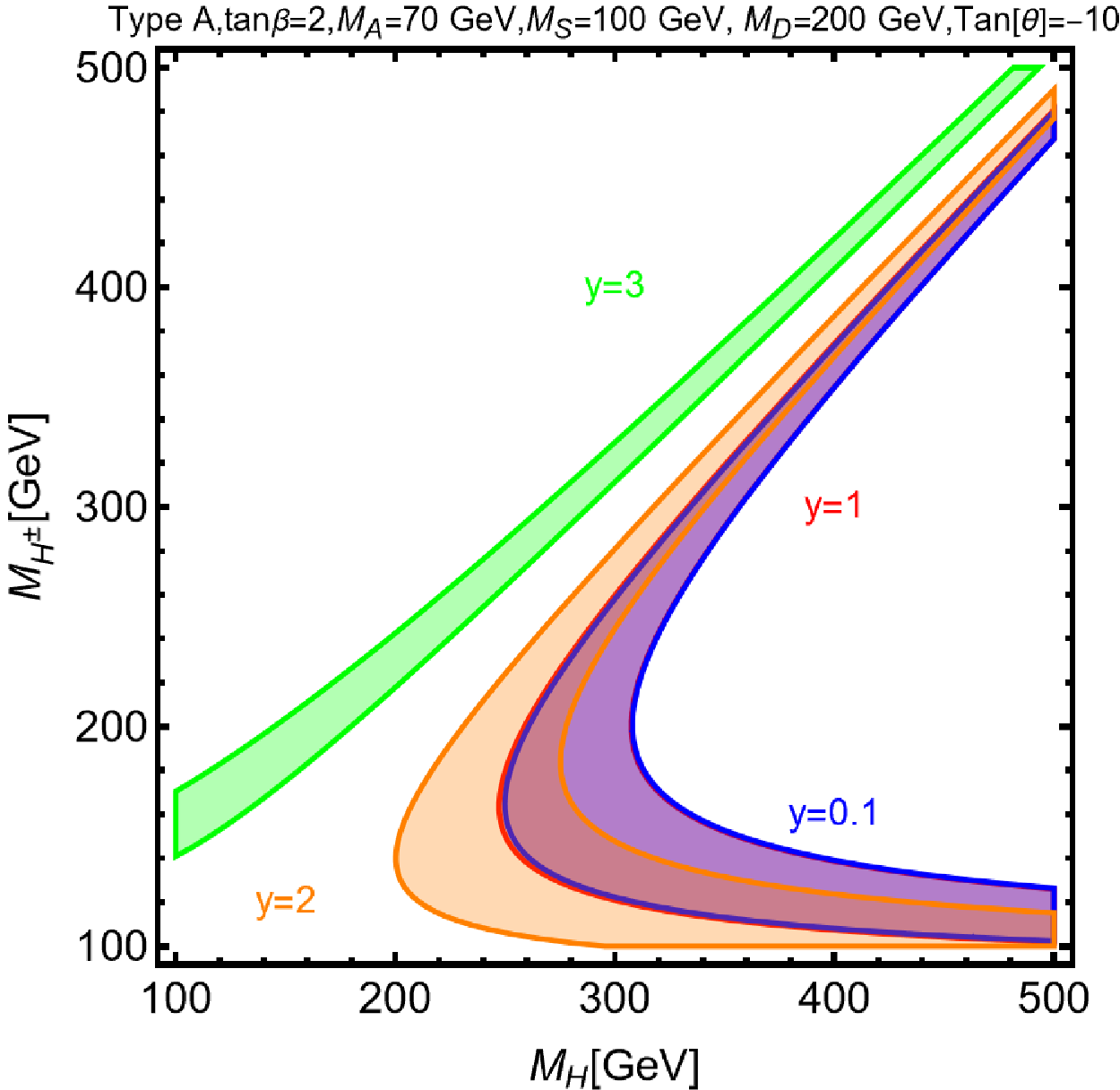}~~~
    \includegraphics[width=.43\linewidth]{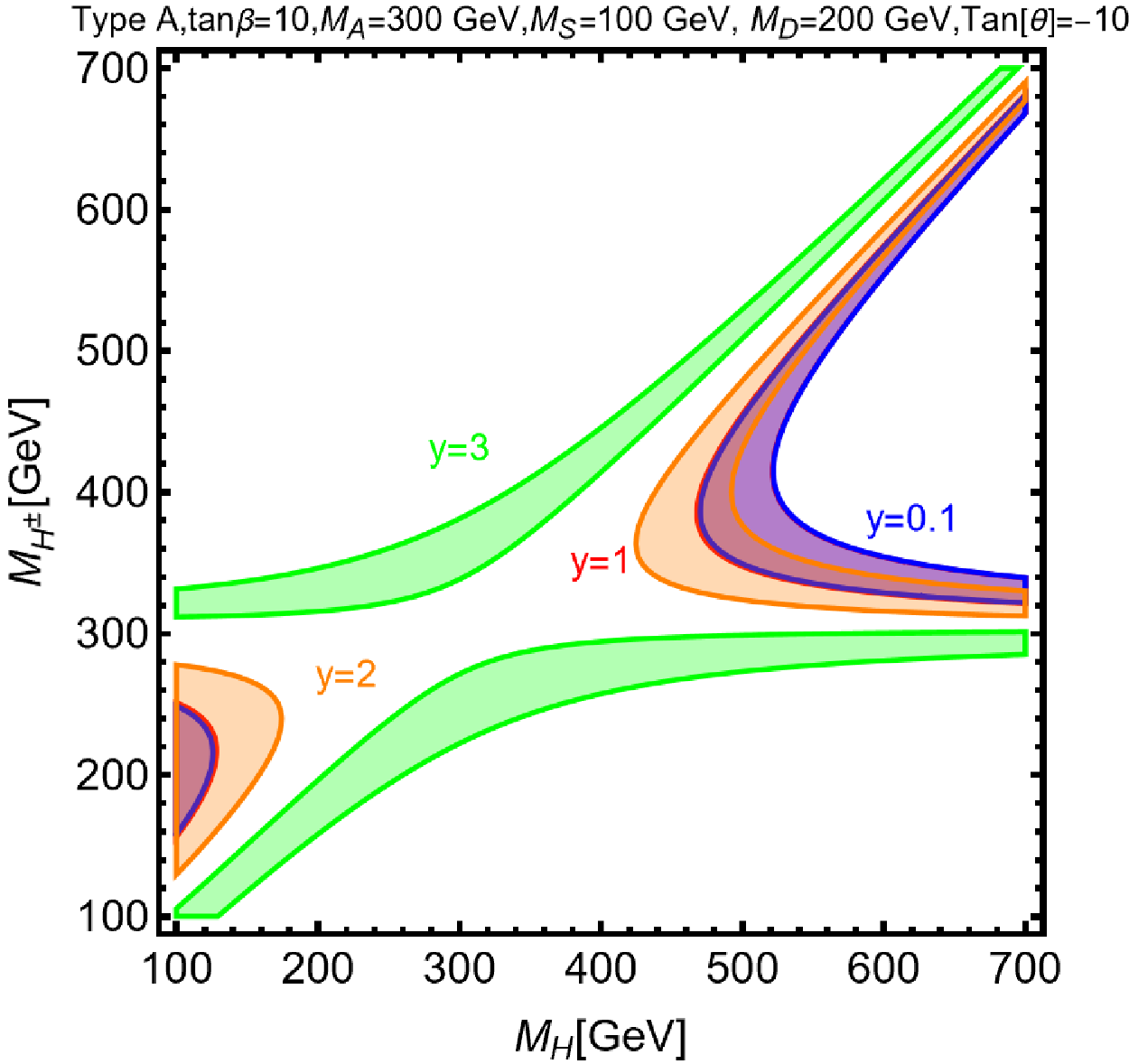}\\[2mm]
     \includegraphics[width=.43\linewidth]{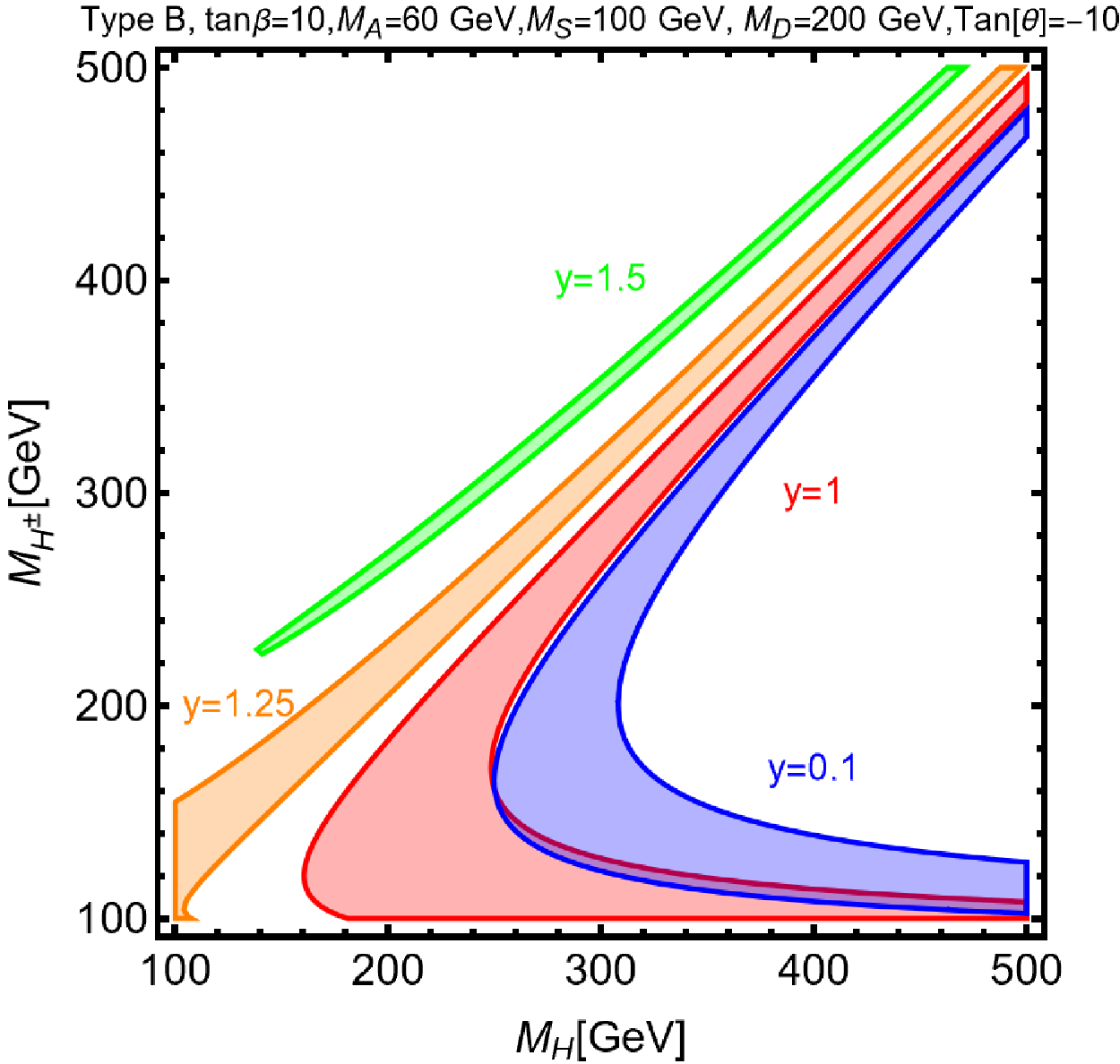}~~~
    \includegraphics[width=.43\linewidth]{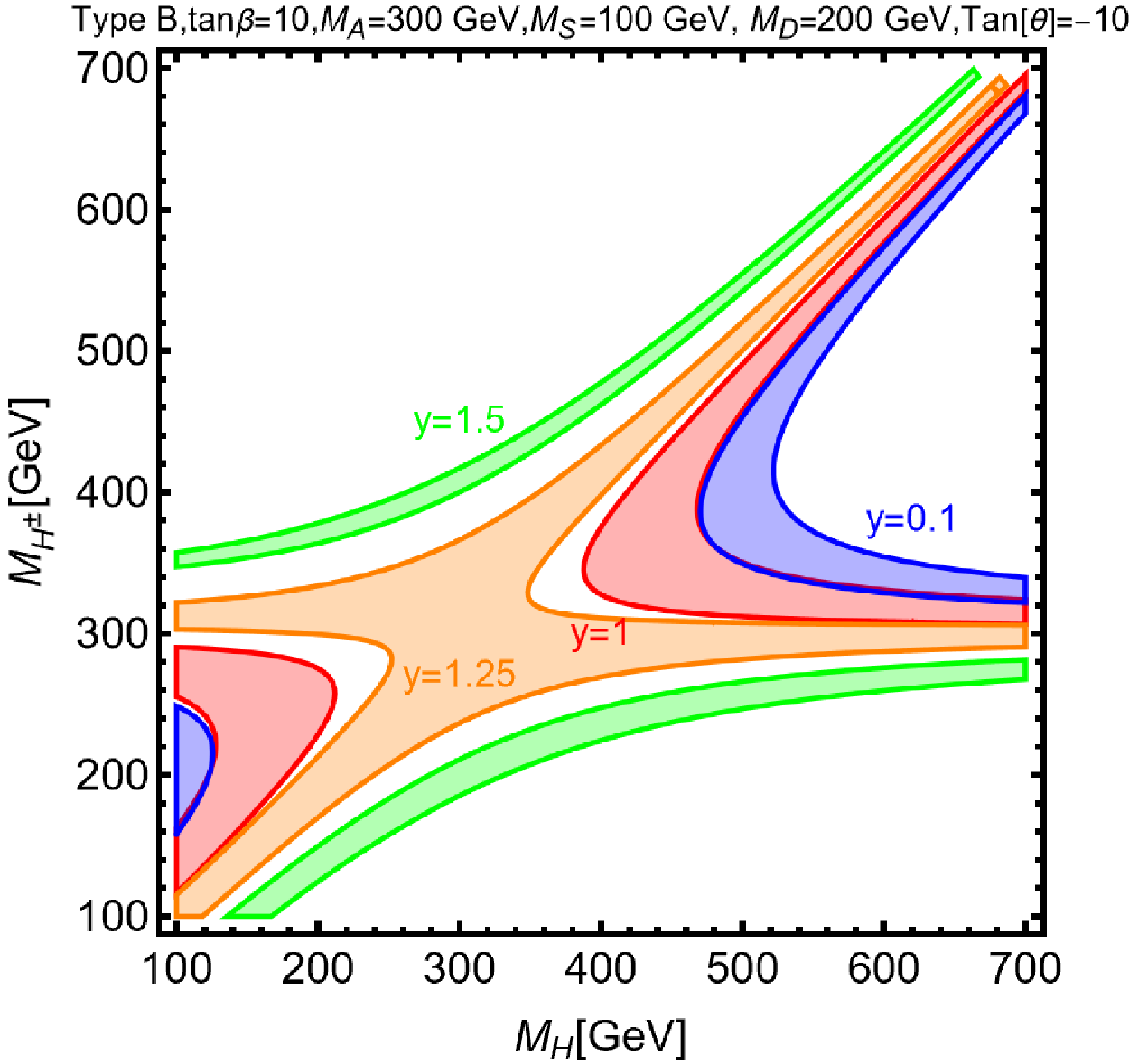}
\vspace*{-2mm}
\caption{Regions in the $[M_H,M_{H^{\pm}}]$ plane providing a viable fit of 
the $M_W^{\rm CDF}$ measurement for some benchmarks of the 2HDM singlet-doublet model. The different colors correspond to the different values of the Yukawa coupling $y$ reported on the panels. {The first row refers to the Type-A configuration of the couplings of the DM while the second row corresponds to the Type-B configuration.}}
    \label{fig:CDF2HDM}
\vspace*{-2mm}
\end{figure}

We show in Fig.~\ref{fig:CDF2HDM} how a viable fit of the $M_W^{\rm CDF}$ anomaly is obtained for two benchmark assignments of the model parameters. The two panels  show the $[M_H,M_{H^{\pm}}]$ plane for two values 
of $M_A$, namely 70 GeV and 300 GeV, with $\tan\beta=10$ in both cases.  

 For what concerns the fermionic sector, we have taken for both panels $m_S=100\,\mbox{GeV}$, $m_D=200\,\mbox{GeV}, \tan\theta=-10$ and considered four values of $y$, corresponding to the different colored contours. We have focused again on a negative value for $\tan\theta$, so that a blind spot can be enforced in DM direct detection.

 {The pattern in the figures can be understood as follows. For the lowest values of the DM Yukawa coupling $y$, the contributions of the new fermions to the electroweak observables and hence to $M_W$ is very small. The $M_W^{\rm CDF}$ result is accounted for mainly by the scalar sector. In agreement with the findings of e.g.  Refs.~\cite{Arcadi:2022dmt,Arcadi:2022lpp}, this result is achieved by taking an appropriate mass splitting between the $H$ and $H^{\pm}$ states. When the Yukawa coupling $y$ increases, the impact of the new fermions on $M_W$ is more significant. Consequently, one has to reduce the contribution for the extra Higgs bosons by having a smaller mass splitting. As it should be clear from the parameter assignment, it is necessary to consider small values of $\tan\beta$ in the Type-A scenario (the DM Yukawa couplings are suppressed with $\tan\beta$) and values of $y$ greater than unity to have a substantial contribution from the new fermions to the CDF anomaly. In the Type-B case,   the strength of the DM interactions increases with $\tan\beta$ and,  consequently, lower values of $y$ are required.}

%\textcolor{red}{Maybe here also a sentence or two to explain briefly what is happening in the figure?} 

\subsection{Addressing the muon g-2 anomaly}

In contrast to the minimal singlet-doublet model discussed in the previous section, the presence of an extended Higgs sector allows also to generate an additional contribution to the anomalous magnetic moment of the muon which could potentially reproduce the recent experimental result. Such a contribution actually emerges from the combination of two types of terms. A first one, which appears at the one-loop level, scales as $m_\mu^2/M_\phi^2$ with $\phi$ being an electrically neutral state of the model. Consequently, it is strongly suppressed unless $\phi$ is very light and we will consider such possibility only for the CP-even $A$ boson. The corresponding contributions can be approximately written as \cite{Dedes:2001nx,Djouadi:1989md}
\begin{equation}
    \Delta a_\mu^{\rm 1\!-\!loop} \approx - \frac{ \alpha_{\rm EM}}{8 \pi \sin^2\theta_W}  \frac{ m^4_\mu}{M_W^2 M_A^2} \; g_{A\mu \mu}^2 \; \bigg[ {\rm log} \bigg( \frac{M_A^2}{m_\mu^2} \bigg) - \frac{11}{6} \bigg] \, .
\end{equation}
Given the already mentioned suppression, a proper computation of $\Delta a_\mu$ should include also the two-loop level contribution which arises from Barr-Zee type diagrams \cite{Barr:1990vd} in which there is a heavy fermion loop with an enhanced $m_f^2/M_{\phi}^2$ term that  compensates the higher $\alpha_{\rm EM}$ power suppression. In the case of the $A$ state, it can be written as \cite{Chang:2000ii,Larios:2001ma,Ilisie:2015tra}
\begin{eqnarray}
\Delta a_\mu^{\rm 2\!-\!loop} = \frac{\alpha_{\rm EM}^2 }{8 \pi^2 \sin^2\theta_W} \; \frac{m_\mu^2}{M_W^2} g_{A\mu \mu}\; \sum_f g_{Aff} N_c^f Q_f \; \frac{m_f^2}{M_A^2} \; H \bigg( \frac{m_f^2}{M_A^2} \bigg) \, , \\[-3mm] 
H(r) = \int_0^1 {\rm d}x \frac{ \log (r)- \log [x(1-x)] } {r -x(1-r) } \, . \hspace*{3cm}
%\end{equation}
\end{eqnarray}
Our numerical determination of $\Delta a_\mu$ is nonetheless obtained by considering the full computation, as given for example in Ref.~\cite{Ilisie:2015tra}, which includes the contribution of all Higgs bosons of the 2HDM. It is arguable from the expressions above that sizable couplings of the new Higgs bosons with the muons are needed to account for the $(g-2)_\mu$ anomaly; for a more detailed discussion, see for example Ref.~\cite{Arcadi:2021zdk}. This requirement selects the Type-II and the Type-X among the flavor preserving Yukawa configurations, as they involve enhanced Higgs couplings to muons at high $\tan\beta$ values, $g_{A\ell\ell} \propto \tan\beta$. However, in the Type-II scenario, the presence of light neutral Higgs bosons is disfavored by  direct Higgs searches at the LHC,  in particular in the production processes $pp \rightarrow gg/b \bar b \to H/A$ and the subsequent decays $H/A \to \tau^+ \tau^-$ \cite{ATLAS:2020zms,CMS:2022rbd}  and eventually also $H/A \to \mu^+\mu^-$ \cite{CMS:2019buh,LHCb:2020ysn}; see again the recent analyses performed in Refs.~\cite{Argyropoulos:2022ezr,Arcadi:2021zdk,Arcadi:2022dmt,Arcadi:2022lpp}.  

\begin{figure}[!ht]
\vspace*{-2mm}
    \centering \mbox{
    \includegraphics[width=0.42\linewidth]{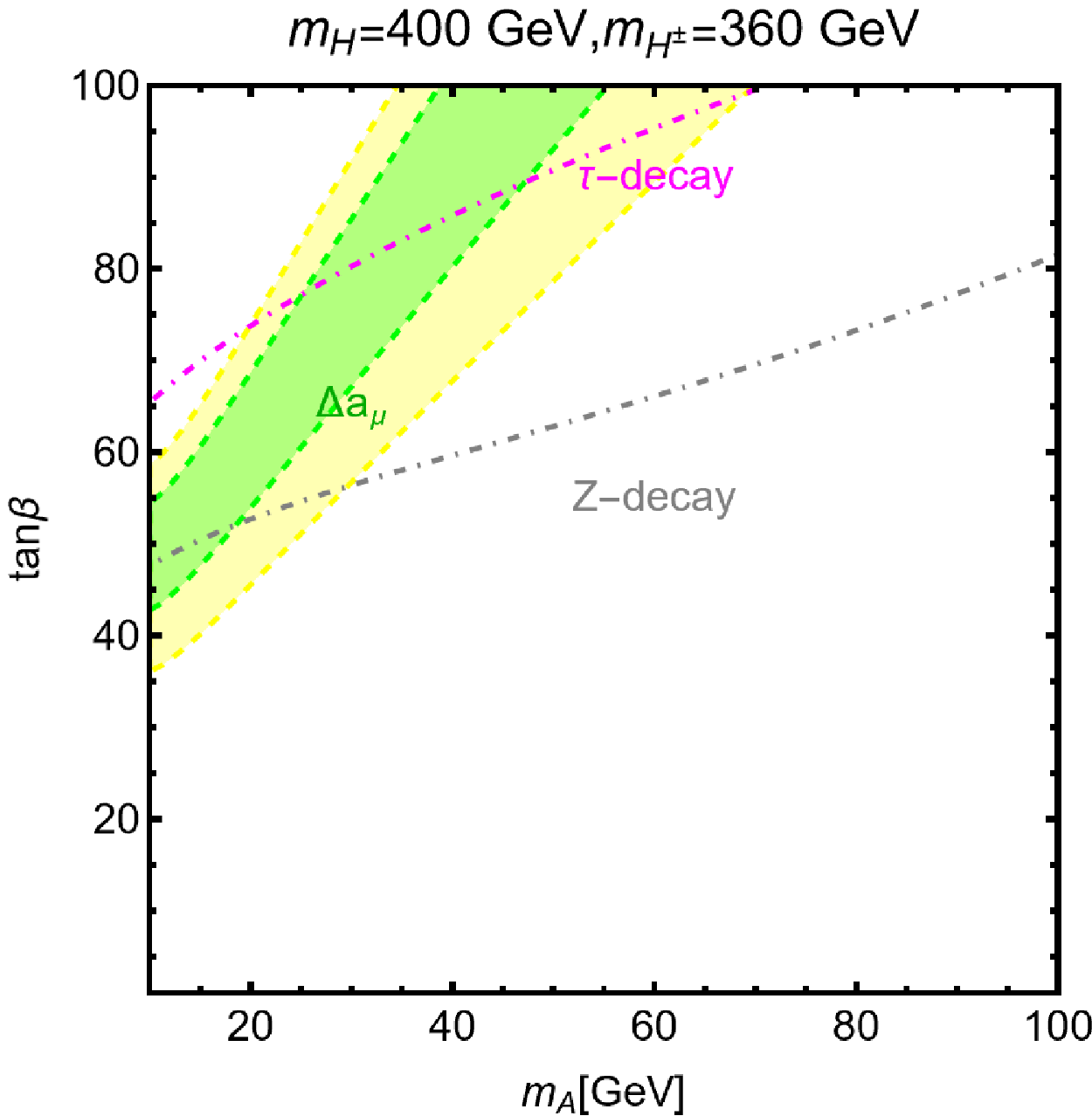}~~
    \includegraphics[width=0.42\linewidth]{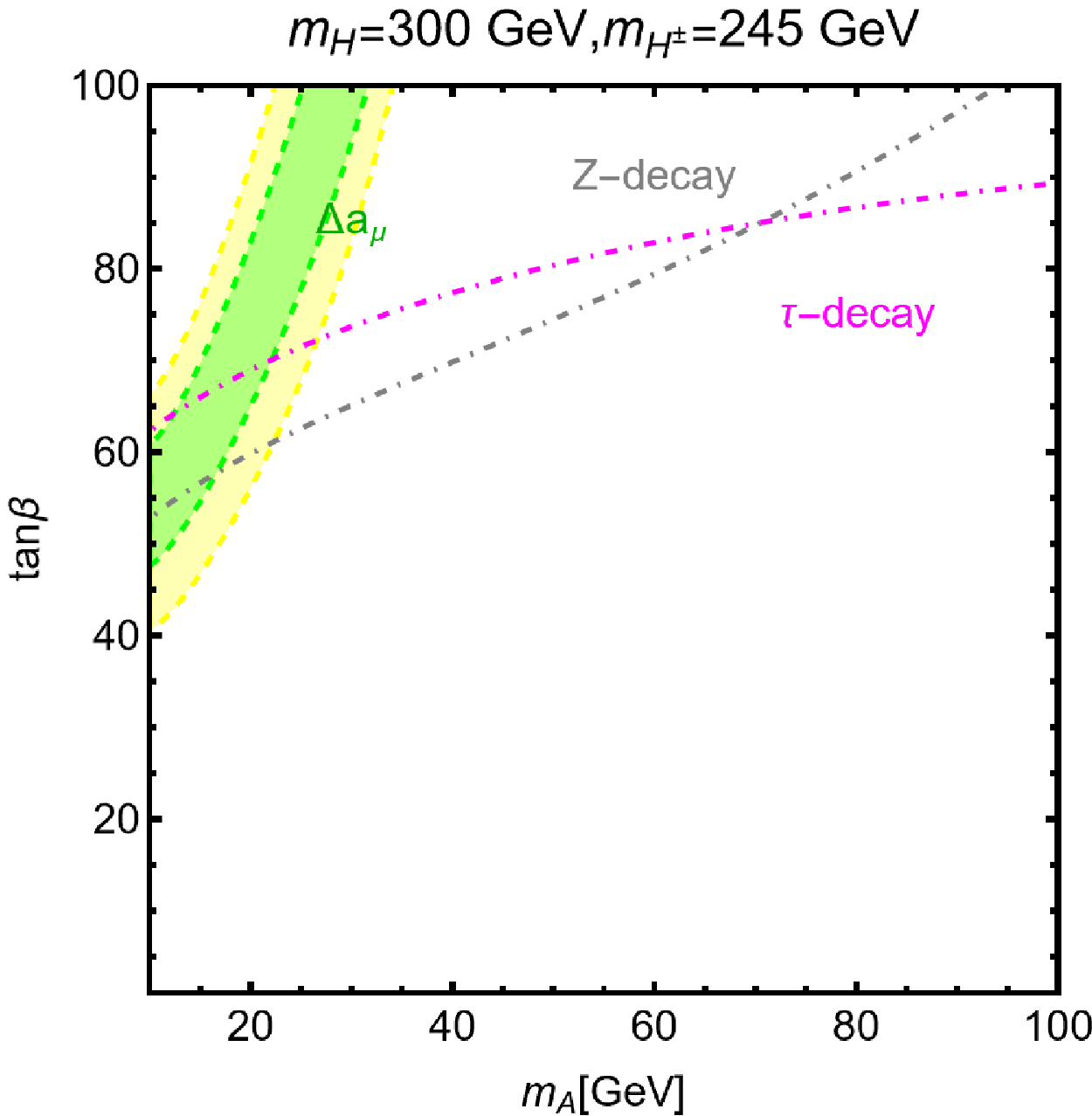}}
    \vspace*{-2mm}
    \caption{Regions providing a viable fit of the $(g\!-\!2)_\mu$ anomaly at $1\sigma$(green) and $2\sigma$ (yellow) in the plane $[M_A,\tan\beta]$ for two assignments of $(M_H,M_{H^{\pm}})$, namely $(400,350)$ and $(300,245)\,\mbox{GeV}$. The dot-dashed lines represent the bounds from violation of lepton universality in decays of the $Z$ boson and $\tau$-lepton (regions above the lines are excluded).}
    \label{fig:plotg2}
\vspace*{-3mm}
\end{figure}

We show in Fig.~\ref{fig:plotg2} the regions of the $[M_A,\tan\beta]$ plane which provide a viable fit of the $(g\!-\!2)_\mu$ anomaly at the $1\sigma$ (green) and $2\sigma$ (yellow) levels. The two panels differ by the assignments of the $(M_H,M_{H^{\pm}})$ pair, which have been chosen, in agreement with the outcome of Fig.~\ref{fig:CDF2HDM}, to provide also a good fit of $M_W^{\rm CDF}$. Furthermore, to overcome the constraint from the $h$ width at low $M_A$ values, we have fixed the value of the angle $\alpha$ as in eq.~(\ref{eq:blindspot}). As can be seen, a viable fit of the $(g-2)_\mu$ anomaly is achieved for very high $\tan\beta$ values and $M_A \lesssim 60\,\mbox{GeV}$.  The sizable mass splitting between the pseudoscalar state $A$ and the other 2HDM states is also constrained by violation of lepton universality in decays of the SM $Z$ and $\tau$ particles \cite{Chun:2016hzs}. The corresponding bounds are represented as dot-dashed isocontours in the figure and the regions above the contours are ruled-out. 

{The different shapes of the $(g-2)_\mu$ contours can be explained as follows. The value of $a_\mu$ in the 2HDM is due to a non-trivial interplay between 1- and 2-loop contributions, as the latter one can potentially exceed the former since the suppression by the factor $\alpha_{\rm EM}$ is compensated by an ${m_f^2}/{m_\mu^2}$ enhancement for $m_f \! \gg\! m_\mu$. Note that the $A$ boson gives a negative (positive) contribution a 1-loop (2-loop) while the opposite occurs for the CP-even $h,H$ bosons. Thus, a good fit of  $(g-2)_\mu$ is obtained with the 2-loop $A$ contribution. In the Type-X case,  this occurs for ${\mathcal O}(10\,\mbox{GeV}) \! <\! M_A \! < \! M_h$ and very high $\tan\beta$ values. The $H$ state should be heavy enough for its negative 2-loop contribution to be reduced. The two panels of Fig. \ref{fig:plotg2} differ also because of the different value of the angle $\alpha$ obtained from  the condition eq.~(\ref{eq:blindspot})
which modifies the couplings of the CP-even $h,H$ states.}\vspace*{-1mm}

{Note that the constraints from $Z$-decays are stronger when the hierarchy between $M_A$ and $M_H$ increases, while constraints from $\tau$ decays become weaker with increasing $M_{H^\pm}$ and have more impact in the right panel. In all cases, these constraints are strong, reducing to a narrow strip the regions in which the $(g-2)_\mu$ value can be reproduced at the $1\sigma$ level.}\vspace*{-1mm}

\subsection{Combined results}

\begin{figure}[!ht]
    \centering \mbox{
    \includegraphics[width=0.43\linewidth]{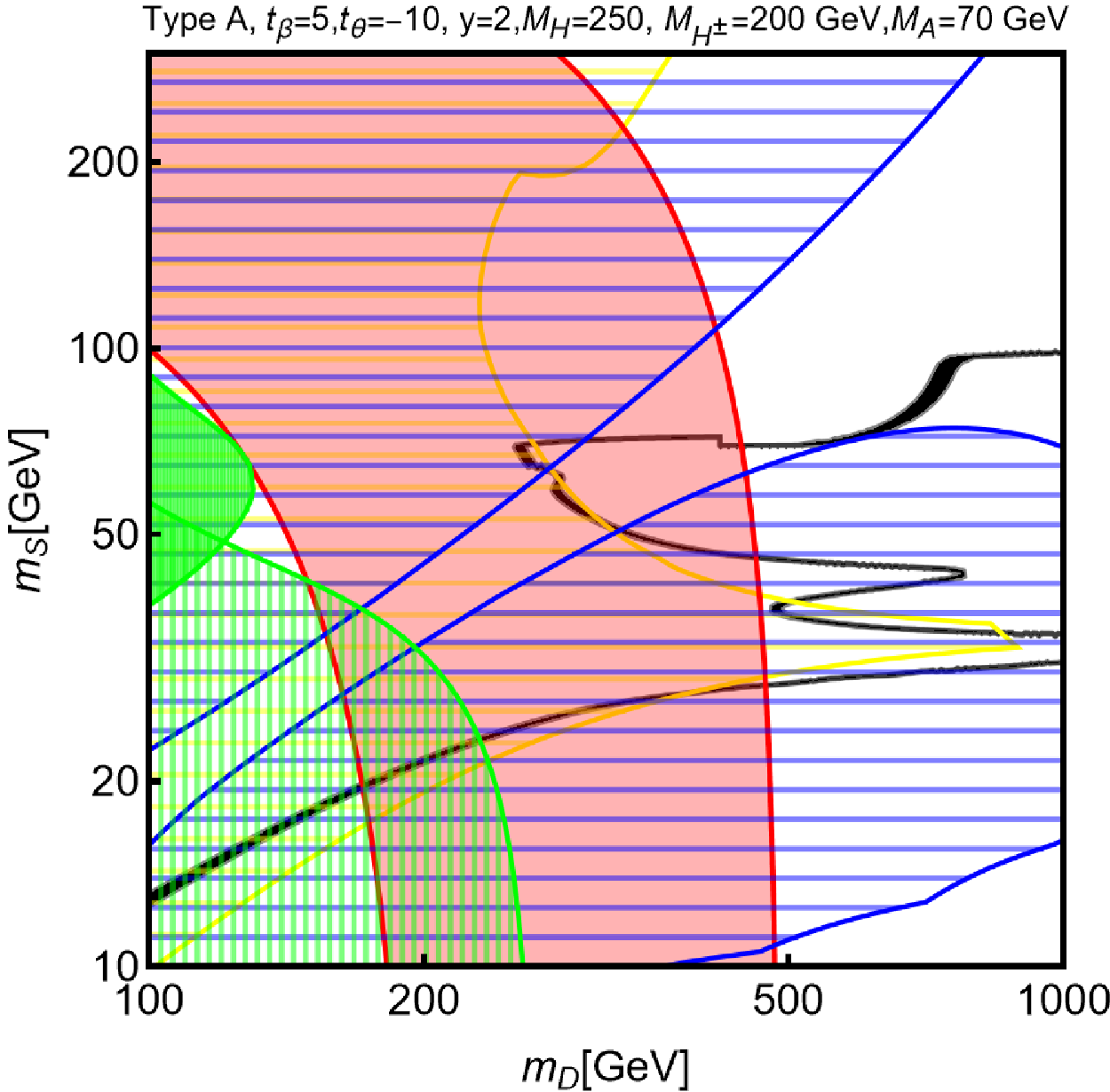}~~~
    \includegraphics[width=0.43\linewidth]{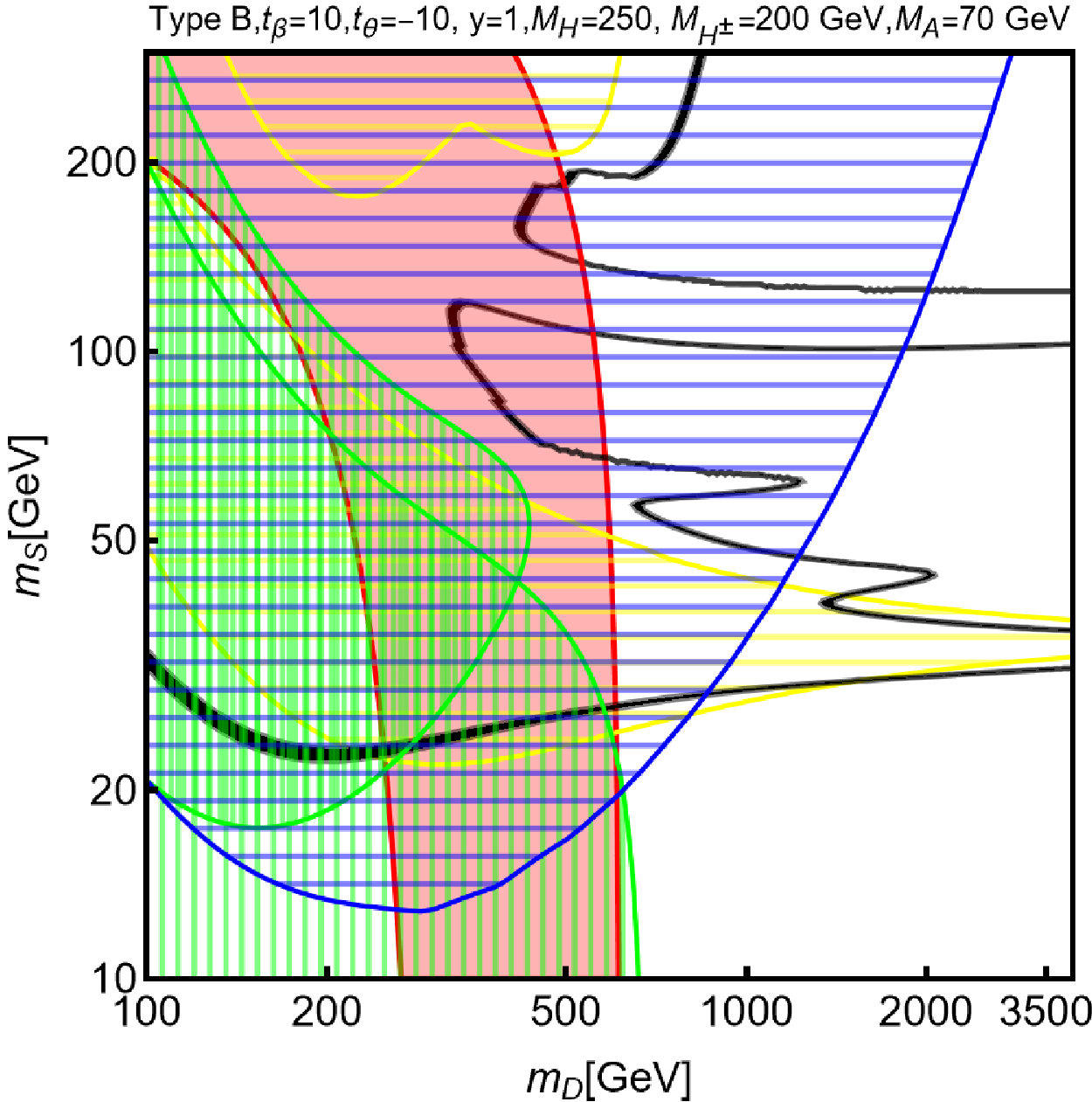}}\\[2mm]
    \includegraphics[width=0.43\linewidth]{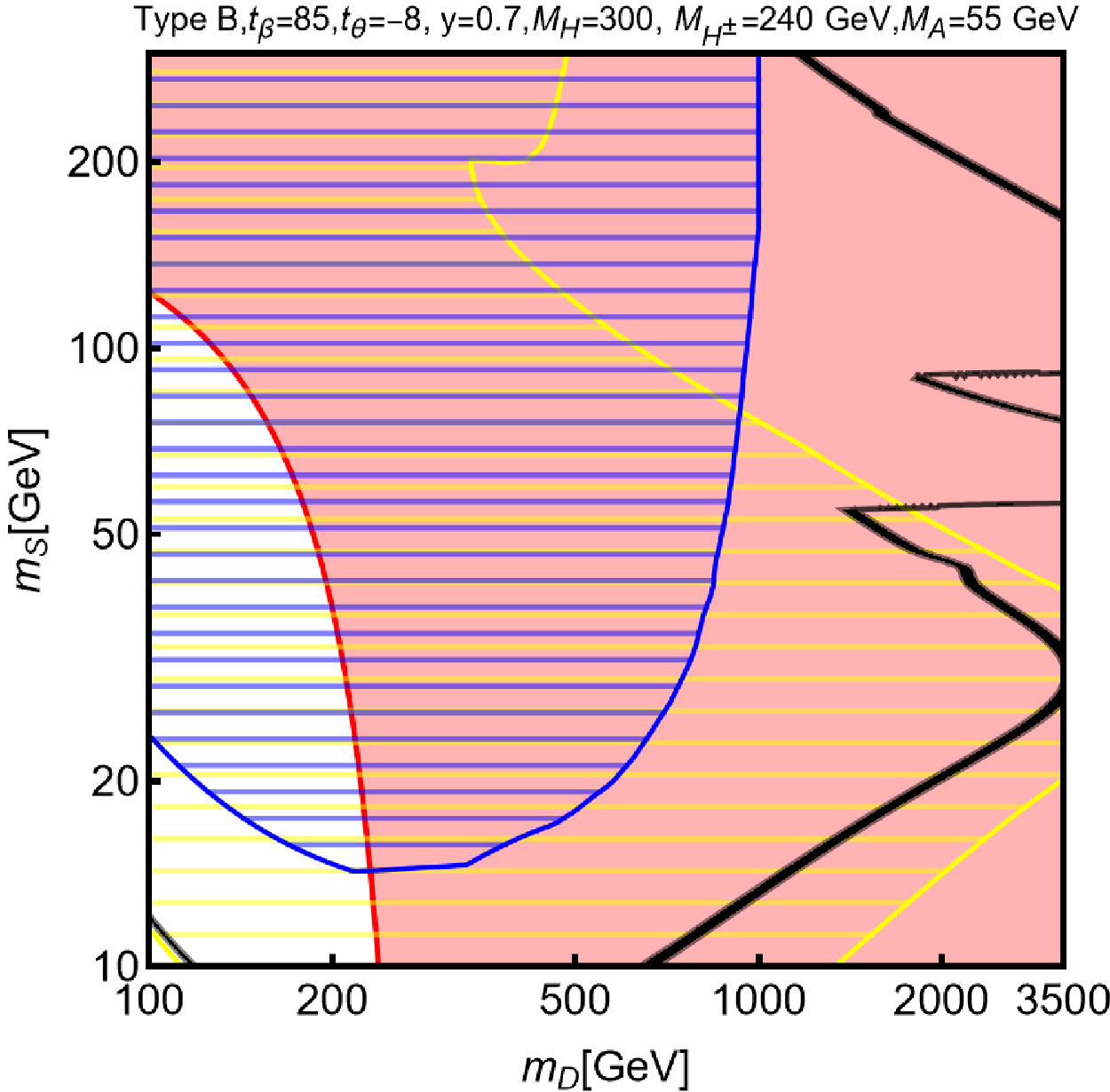}
\vspace*{-1mm}
    \caption{Combined constraints on the singlet-double 2HDM scenario in the 
     $[m_D,m_S]$ plane for three benchmarks in the Type-X configuration, with input parameters given on top of each plot. The color convention is the same as in Fig.~\ref{fig:plotSD}.}
    \label{fig:plot2HDM}
\vspace*{-5mm}
\end{figure}

We are now again ready to combine the individual constraints previously discussed to obtain the global picture that is shown in Fig.~\ref{fig:plot2HDM}. We have considered three benchmarks for the singlet-doublet 2HDM model in the Type-X configuration and with the $\tan\beta$ and $M_\phi$ values listed on each plot, and imposed the various constraints in the $[m_D,m_S]$ plane. The color code is the same as the one adopted in the minimal singlet-doublet case, Fig.~\ref{fig:plotSD}. A combined fit of the correct relic density and of the $M_W^{\rm CDF}$ result is achieved at the crossing of the black line (relic density) and red area ($M_W$). Such an intersection should lie outside the experimental exclusions, represented by the hatched colored regions.

In the three selected benchmarks,  the two first ones allow simply to combine the CDF result for $M_W$ with DM phenomenology and similar parameter assignments as in Fig.~\ref{fig:pmwSD} of the minimal singlet-doublet model have been adopted. Comparing the outcome with the analogous one given in the previous section, one first notices that the relic density contours have a richer pattern. This is due to the presence of the possibly  light extra Higgs bosons which could meet the resonance condition, $m_{\chi_1} \simeq \frac12 M_{\phi}$ for the relic density.\vspace*{-1mm} 

A second notable difference with respect to the minimal model is that, due to the dependence of the entries of the Majoana mixing matrix $U$ on $\tan\beta$, it is possible to further reduce  the impact of the bounds from DM direct detection and from the invisible Higgs decay branching ratio, in addition to the blind spot condition for negative $\tan\theta$ values. Notice that for these first two benchmarks, we have considered only values $M_A > \frac12 M_h$, so that the ``invisible" width of the $h$ state is due only to the decay into DM pairs. In the figures, in  addition to the limits from direct detection \cite{LZ:2022ufs,XENON:2019rxp}, we have also explicitly shown the parameter space excluded by DM indirect detection, represented by the negative results of the searches of $\gamma$-ray signals performed by the FERMI-LAT experiment \cite{Fermi-LAT:2015rbk,Fermi-LAT:2015kyq}.

In turn, the third benchmark of  Fig.~\ref{fig:plot2HDM}  is characterized by a very high value of $\tan\beta$ and a light $A$ boson, $\tan\beta=85$ (which allows a perturbativity of all couplings in the Type-X case) and $M_A=55$ GeV (which is not excluded by $pp \! \to \! A \! \to \tau\tau, \mu\mu$ searches). As can be seen from the last panel of Fig.~\ref{fig:plot2HDM},  this benchmark leads at the same time to a correct DM relic density and provides viable interpretations of both the $(g-2)_\mu$  and $M_W^{\rm CDF}$ anomalies.

\section{Conclusions}

In this work, we have considered the relatively simple fermionic singlet-doublet model for dark matter, first with a minimal Higgs sector and then with an extended one to include two doublets.  We have explored the possibility of simultaneously fulfilling the collider and astroparticle physics constraints that allow to obtain a successful DM candidate with the correct relic density, and addressing two recent experimental anomalies, namely the discrepancies with respect to the prediction in the SM of the muon anomalous magnetic moment $(g\!-\!2)_\mu$ and the mass of the $W$ boson $M_W$ measured by the CDF collaboration. 

We have shown that in the minimal singlet-doublet model with a SM-like Higgs sector, as a result of the presence of a new fermionic sector coupled with the SM gauge bosons, one can address only the $M_W^{\rm CDF}$ anomaly while having a DM with the correct relic density. The extra particle spectrum does not couple to SM fermions and cannot explain the experimental $(g\!-\!2)_\mu$ value.  Nevertheless, the model parameter space is almost entirely excluded by the constraints on the DM particle that arise from direct detection. 

Extending the Higgs sector of the singlet-doublet model to contain a second scalar doublet field is doubly beneficial. On the one hand, it allows to evade the constraints from DM direct detection and, on the other hand, one can also achieve a viable interpretation of the muon $(g-2)$ anomaly, besides the interpretation of the CDF $M_W$ measurement. This is done by means of a light pseudoscalar $A$ boson that strongly couples to muons. In this case, significant parts of the parameter space of the model are still allowed but they will be challenged by the next round of collider and astroparticle physics experiments.\smallskip

\noindent {\bf Acknowledgements:} 
AD is supported by the Estonian Research Council (ERC) grant MOBTT86 and by the Junta de Andalucia through the Talentia Senior program and the grants PID2021-128396NB-I00, A-FQM-211-UGR18 and P18-FR-4314 with ERDF.

\bibliographystyle{frontiersinHLTH&FPHY} % for Physics and Mathematics articles
\bibliography{biblio}
\end{document}